\documentclass[12pt,letterpaper]{JHEP3}
\usepackage{amsmath}
\usepackage{amsthm}
\usepackage{amssymb}
\usepackage{cite}
\usepackage{graphicx}
\usepackage{subfigure}

\newcommand{\comment}[1]{}
\newcommand{\abs}[1]{\left| #1\right|}  

\title{Star Formation in the Multiverse}

\author{Raphael Bousso and Stefan Leichenauer\\
  Center for Theoretical Physics, Department of Physics\\
  University of California, Berkeley, CA 94720-7300, U.S.A.\\
  {\em and}\\
  Lawrence Berkeley National Laboratory, Berkeley, CA 94720-8162,
  U.S.A.}

\abstract{We develop a simple semi-analytic model of the star
  formation rate (SFR) as a function of time.  We estimate the SFR for
  a wide range of values of the cosmological constant, spatial
  curvature, and primordial density contrast.  Our model can predict
  such parameters in the multiverse, if the underlying theory
  landscape and the cosmological measure are known.}
  
\begin{document}

\section{Introduction}

In this paper, we study cosmological models that differ from our own
in the value of one or more of the following parameters: The
cosmological constant, the spatial curvature, and the strength of
primordial density perturbations.  We will ask how these parameters
affect the rate of star formation (SFR), i.e., the stellar mass
produced per unit time, as a function of cosmological time.

There are two reasons why one might model the SFR.  One, which is not
our purpose here, might be to {\em fit\/} cosmological or
astrophysical parameters, adjusting them until the model matches the
observed SFR.  Our goal is different: We would like to {\em explain\/}
the observed values of parameters, at least at an order-of-magnitude
level.  For this purpose, we ask what the SFR would look like if some
cosmological parameters took different values, by amounts far larger
than their observational uncertainty.

This question is reasonable if the observed universe is only a small
part of a multiverse, in which such parameters can vary.  This
situation arises when there are multiple long-lived vacua, such as in
the landscape of string theory.  It has the potential to explain the
smallness of the observed cosmological constant~\cite{BP}, as well as
other unnatural coincidences in the Standard Model and in Standard
Cosmology.  Moreover, a model of star formation can be used to test
the landscape, by allowing us to estimate the number of observers that
will observe particular values of parameters.  If only a very small
fraction of observers measure the values we have observed, the theory
is ruled out.

The SFR can be related to observers in a number of ways.  It can serve
as a starting point for computing the rate of entropy production in
the universe, a well-defined quantity that appears to correlate well
with the formation of complex structures such as
observers~\cite{Bou06}.  (In our own universe, most entropy is
produced by dust heated by starlight~\cite{BouHar07}.)  Alternatively,
one can posit that a fixed (presumably very small) average number of
observers arises per star, perhaps after a suitable delay time of
order billions of years.

To predict a probability distribution for some parameter, this
anthropic weighting must be combined with the statistical distribution
of parameters in the multiverse.  Computing this distribution requires
at least statistical knowledge of the theory landscape, as well as
understanding the measure that regulates the infinities arising in the
eternally inflating multiverse.  Here, we consider none of these
additional questions and focus only on the star formation rate.

We develop our model in Sec.~\ref{sec-model}.  In
Sec.~\ref{sec-examples} we display examples of star formation rates we
have computed for universes with different values of curvature,
cosmological constant, and/or primordial density contrast.  We
identify important qualitative features that emerge in exotic regimes.

\section{Model}
\label{sec-model}

\subsection{Goals}

Our goal is not a detailed analytical fit to simulations or
observations of the SFR in our universe.  Superior models are already
available for this purpose.  For example, the analytic model of
Hernquist and Springel~\cite{HS} (HS) contains some free parameters,
tuned to yield good agreement with simulations and data. However, the
HS model only allows for moderate variation in cosmological
parameters.  For example, HS estimated the SFR in cosmologies with
different primordial density contrast, $Q$, but only by 10\%.

Here we are interested in much larger variations, by orders of
magnitude, and not only in $Q$ but also in the cosmological constant,
$\Lambda$, and in spatial curvature. A straightforward extrapolation
of the HS model would lead to unphysical predictions in some parameter
ranges. For example, a larger cosmological constant would actually
enhance the star formation rate at all
times~\cite{Cline}.\footnote{This effect, while probably
  unphysical, did not significantly affect the probability
  distributions computed by Cline {\em et al.}~\cite{Cline}.}

In fact, important assumptions and approximations in the HS model
become invalid under large variations of parameters.  HS assume a
fixed ratio between virial and background density.  But in fact, the
overdensity of freshly formed haloes drops from about 200 to about 16
after the cosmological constant dominates~\cite{Tegmark}.  Moreover,
HS apply a ``fresh-halo approximation'', ascribing a virial density
and star formation rate to each halo as if it had just been formed.
However, large curvature, or a large cosmological constant, will
disrupt the merging of haloes.  One is left with an old population of
haloes whose density was set at a much earlier time, and whose star
formation may have ceased due to feedback or lack of baryons.
Finally, HS neglect Compton cooling of virialized haloes against the
cosmic microwave background; this effect becomes important if
structure forms earlier, for example because of increased
$Q$~\cite{TegmarkRees}.

All these effects are negligible at positive redshift in our universe;
they only begin to be important near the present era.  Hence, there
was no compelling reason to include them in conventional,
phenomenologically motivated star formation models.  But for different
values of cosmological parameters, or when we study the entire
evolution of the universe, these effects become important and must be
modelled.

On the other hand, some aspects of star formation received careful
attention in the HS model in order to optimize agreement with
simulations.  Such refinements are unnecessary when we work with huge
parameter ranges, where uncertainties are large in any case.  At best,
our goal can be to gain a basic understanding of the quantitative
behavior of the SFR as parameters are varied.  We will mainly be
interested in the height, time, and width of the peak of the SFR
depend on parameters.  In currently favored models of the
multiverse~\cite{Bou06,DeSimoneetal}, more detailed aspects of the
curve do not play an important role.  Thus, our model will be
relatively crude, even as we include qualitatively new phenomena.

Many extensions of our model are possible.  The most obvious is to
allow additional parameters to vary.  Another is to refine the
treatment of star formation itself.  Among the many approximations and
simplifications we make, the most significant may be our neglect of
feedback effects from newly formed stars and from active galactic
nuclei.  In addition, there are two regimes in our model (high $Q$ and negative $\Lambda$ near the Big Crunch) where we use physical arguments to cut off star formation by hand.  These are regimes where the star formation rate is so fast that neglected timescales, such as halo merger rates and star lifetimes, become important.  An extended model which accounts for these timescales may avoid the need for an explicit cutoff.

\subsection{Geometry and initial conditions}

Our star formation model can be applied to open, flat, and closed
Friedman-Robertson-Walker (FRW) universes.  However, we will consider
only open geometries here, with metric
\begin{equation}
ds^2=-dt^2+a(t)^2\left(d\chi^2 + \sinh^2\chi d\Omega^2\right)~.
\end{equation}
Pocket universes produced by the decay of a false vacuum are open FRW
universes~\cite{CDL}; thus, this is the case most relevant to a
landscape of metastable vacua.  The above metric includes the flat FRW
case in the limit where the spatial curvature radius $a$ is much
larger than the horizon:
\begin{equation}
a \gg H^{-1}~,
\end{equation}
where $H\equiv \dot{a}/a$ is the Hubble parameter.  This can be
satisfied at arbitrarily late times $t$, given sufficiently many
$e$-foldings of slow roll inflation.

The scale factor, $a(t)$, can be computed by integrating the Friedmann
equations,
\begin{eqnarray} 
  H^2-\frac{1}{a^2} & = &
  \frac{8\pi G_{\rm N}}{3}~\rho~,\\
  \dot{H} + H^2 & = & -\frac{4\pi G_{\rm N}}{3}(\rho+3p)~,
\end{eqnarray} 
beginning at (matter-radiation) equality, the time $t_{\rm eq}$ at which matter and radiation have equal density:
\begin{equation}
  \rho_{\rm m}(t_{\rm eq}) =  
  \rho_{\rm r}(t_{\rm eq})\equiv \nu\frac{\pi^2}{15} T_{\rm eq}^4~,
  \label{eq-mr}
\end{equation}
where
\begin{equation} 
\nu=1+\frac{21}{8}\left(\frac{4}{11}\right)^{4/3}\approx 1.681
\end{equation}
accounts for three species of massless neutrinos.  The temperature at
equality is~\cite{Tegmark}
\begin{equation}
T_{\rm eq} = 0.22 \xi = 0.82~\mbox{eV}~,
\end{equation}
and $\xi$ is the matter mass per photon.  The time of equality is
\begin{equation} 
  t_{\rm eq}=\frac{\sqrt{2}-1}{\sqrt{3\pi
    G_{\rm N}\rho_{\rm m}(t_{\rm eq})}}\approx 0.128\,G_{\rm N}^{-1/2}\,T_{\rm eq}^{-2} = 4.9 \times 10^4~\mbox{yrs}~.
\end{equation}
The total density and pressure are
given by 
\begin{eqnarray}
  \rho & = & \rho_{\rm r} +\rho_{\rm m} +\rho_\Lambda~,\\
  p & = & \frac{\rho_{\rm r}}{3} - \rho_\Lambda~.  
\end{eqnarray}

Among the set of initial data specified at equality, we will treat
three elements as variable parameters, which define an ensemble of
universes. The first two parameters are the densities associated with
curvature, $G_{\rm N}^{-1} a(t_{\rm eq})^{-2}$, and with the vacuum,
$\rho_\Lambda\equiv\Lambda/8\pi G_{\rm N} $, at equality. We will only
consider values that are small compared to the radiation and matter
density at equality, Eq.~(\ref{eq-mr}); otherwise, there would be no
matter dominated era and no star formation. The third parameter is the
primordial density contrast, $Q$, discussed in more detail in the next
section.

We find it intuitive to trade the curvature parameter $a(t_{\rm eq})$
for an equivalent parameter $\Delta N$, defined as
\begin{equation}
\Delta N = \log\left[\frac{a(t_{\rm eq})}{a_{\rm min}(t_{\rm eq})}\right]~,
\end{equation}
where 
\begin{equation}
  a_{\rm min}(t_{\rm eq}) = 
  \left(\frac{45\, \Omega_{\rm m}}{\nu~8\pi^3\,G_{\rm N}\,T_{\rm eq}^4\,
      H_0}\right)^{1/3}(\Omega_k^{\rm max})^{-1/2}=1.13\times10^{24}\,\mbox{m} = 37\,\mbox{Mpc}~,
\end{equation}
is the experimental lower bound on the curvature radius of {\em our\/}
universe at matter-radiation equality. (Consistency with our earlier
assumption of an open universe requires that we use the upper bound on
{\em negative\/} spatial curvature, i.e., on positive values of
$\Omega_k$.  We are using the 68\% limit from
WMAP5+BAO+SN~\cite{Komatsu}, $\Omega_k=-0.0050^{+0.0061}_{-0.0060}$.
We use the best fit values $H_0=68.8~\mbox{km/s/Mpc}$ and $\Omega_{\rm
  m}=0.282$ for the current Hubble parameter and matter density
fraction.)

$\Delta N$ can be interpreted as the number of $e$-folds of inflation
in excess (or short) of the number required to make our universe (here
assumed open) just flat enough ($\Omega^{\rm max}_{k}=0.0011$) for its
spatial curvature to have escaped detection.  Requiring that curvature
be negligible at equality constrains $\Delta N$ to the range $\Delta
N> -5$.  Values near this cutoff are of no physical interest in any
case: Because too much curvature disrupts galaxy formation, we will
find that star formation already becomes totally ineffective below
some larger value of $\Delta N$.

We hold fixed all other parameters, including Standard Model
parameters, the baryon to dark matter ratio, and the temperature at
equality. We leave the variation of such parameters to future work.

\subsection{Linear perturbation growth and halo formation}

Cosmological perturbations are usually specified by a time-dependent
power spectrum, $\mathcal{P}$, which is a function of the wavenumber
of the perturbation, $k$.  The r.m.s.\ fluctuation amplitude,
$\sigma$, within a sphere of radius $R$, is defined by smoothing the
power spectrum with respect to an appropriate window function:
\begin{equation}
  \sigma^2 = \frac{1}{2\pi^2}\int_0^\infty \left(\frac{3\sin(kR)-3\,kR\cos(kR)}{(kR)^3}\right)^2
  \mathcal{P}(k)k^2\,dk~.
\end{equation}
The radius $R$ can be exchanged for the mass $M$ of the perturbation,
using $M = 4\pi\rho_{\rm m}R^3/3$.  Starting at $t_{\rm eq}$, once all
relevant modes are in the horizon, the time development of $\sigma$
can be computed using linear perturbation theory.  In the linear
theory, all modes grow at the same rate and so the time dependence of
$\sigma$ is a simple multiplicative factor:
\begin{equation}
\sigma(M,t) = Q\,s(M)\,G(t).
\end{equation}
Here $Q$ is the amplitude of primordial density perturbations, which
in our universe is observed to be $Q_0\approx 2\times 10^{-5}$.
$G(t)$ is the linear growth function, and is found by solving
\begin{equation}
\frac{d^2G}{dt^2}+2H\frac{dG}{dt}=4\pi G_{\rm N}\rho_{\rm m}G
\end{equation}
numerically, with the initial conditions $G=5/2$ and $\dot{G} = 3H/2$ at $t=t_{\rm eq}$.  This normalization is chosen so that $G\approx
1+\frac{3}{2}\frac{a(t)}{a(t_{\rm eq})}$ near $t=t_{\rm eq}$, which is the exact solution for a flat universe consisting only of matter and radiation. For $s(M)$ we
use the fitting formula provided in~\cite{Tegmark}:
\begin{equation}
  s(M) = \left[ 
    \left(9.1\mu^{-2/3}\right)^{-0.27} 
    + 
 \left(50.5\log_{10} \left(834 + \mu^{-1/3}\right)-92\right)^{-0.27}
  \right]^{-1/0.27}
\end{equation}
with $\mu=M/M_{\rm eq} $, where
\begin{equation}
M_{\rm eq} = G_{\rm N}^{-3/2} \xi^{-2} = 1.18\times 10^{17} M_\odot 
\end{equation}
is roughly the mass contained inside the horizon at equality, and
$\xi=3.7$ eV is the matter mass per photon. This fitting formula for
$s(M)$ and our normalization convention for $G(t)$ both make use of
our assumption that curvature and $\rho_\Lambda$ are negligible at
$t_{\rm eq}$.

We use the extended Press-Schechter (EPS) formalism~\cite{LaceyCole}
to estimate the halo mass function and the distribution of halo ages.
In this formalism the fraction, $F$, of matter residing in collapsed
regions with mass less than $M$ at time $t$ is given by
\begin{equation}
F(<M,t) = {\rm Erf}\left(\frac{\delta_{\rm c}}{\sqrt{2} \sigma(M,t)}\right)
\label{eq:PSFraction}
\end{equation}
with $\delta_{\rm c}=1.68$ the critical fluctuation for collapse.
$\delta_{\rm c}$ is weakly dependent on cosmological parameters, but
the variation is small enough to ignore within our approximations.

Given a halo of mass $M$ at time $t$, one can ask for the probability,
$P$, that this halo virialized before time $t_{\rm vir}$.  As
in Ref.~\cite{LaceyCole}, we define the virialization time of the halo as
the earliest time when its largest ancestor halo had a mass greater
than $M/2$.  To find $P$, we first define the function
\begin{equation}
\beta(M_1,t_1,M_2,t_2)={\rm Erfc}\left( \frac{\delta_{\rm c}}{Q\sqrt{2 (s(M_1)^2 - s(M_2)^2)}}\left(\frac{1}{G(t_1)}-\frac{1}{G(t_2)}\right)    \right).
\end{equation}
Then the desired probability is given by
\begin{equation}\label{eq:EPSFunction}
P(<t_{\rm vir},M,t) = \int_{M/2}^M \frac{M}{M_1}\frac{d\beta}{dM_1}(M_1,t_{\rm vir},M,t)  \,dM_1.
\end{equation}

The virialization time further determines the density, $\rho_{\rm vir}$, of the halo.  We will compute the virial density by considering the spherical top-hat collapse of an overdense region of space.  Birkhoff's theorem says that we may consider this region to be part of an independent FRW universe with a cosmological constant identical to that of the background universe.  As such, it evolves according to
\begin{equation}
H^2= \frac{8\pi G_{\rm N} \rho_{\rm m}}{3} + \frac{8\pi G_{\rm N} \rho_\Lambda}{3} - \frac{k}{a^2}= \frac{8\pi G_{\rm N} |\rho_{\Lambda}|}{3}\left(\frac{1}{A^3} \pm 1- \frac{\kappa}{A^2}   \right)
\end{equation}
where $A\equiv\left(\frac{|\rho_\Lambda|}{\rho_{\rm m}}\right)^{1/3}$
and $\kappa\equiv\frac{3}{8\pi G_{\rm N}
  |\rho_{\Lambda}|}\left(\frac{A}{a}\right)^2k$. We neglect the
radiation term, which will be negligible for virialization well after
matter-radiation equality.

Assuming that $\kappa$ and $\rho_\Lambda$ are such that the
perturbation will eventually collapse, the maximum value of $A$ is
obtained by solving the equation $H=0$:
\begin{equation}
\pm A_{\rm max}^3 - \kappa A_{\rm max} + 1 =0.
\end{equation}
The time of collapse, which we identify with the halo virialization time $t_{\rm vir}$, is twice the time to reach $A_{\rm max}$ and is given by
\begin{equation}
t_{\rm vir} = 2\sqrt{\frac{3}{8\pi G_{\rm N} |\rho_\Lambda|}}\int_0^{A_{\rm max}} \frac{dA}{\sqrt{\frac{1}{A} \pm A^2- \kappa}}.
\end{equation}
To obtain the virial density, we follow the usual rule that $\rho_{\rm vir}$ is $8$ times the matter density at turnaround.  In other words,
\begin{equation}
\rho_{\rm vir} = 8 \frac{|\rho_\Lambda|}{A_{\rm max}^3}.
\end{equation}

\begin{figure}
\center
\includegraphics[width=3.5 in]{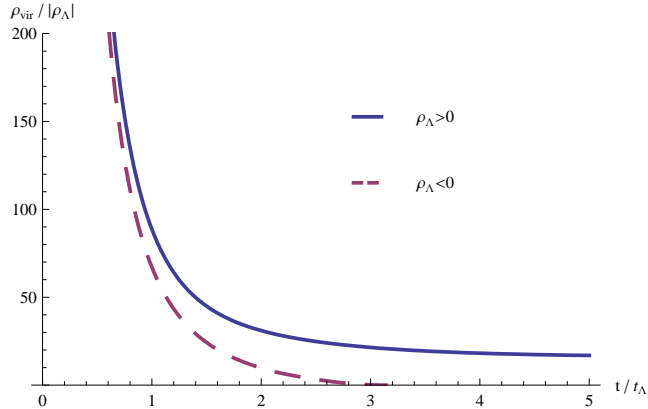}
\caption{The virial density is independent of curvature and has a
  simple scaling dependence on $\rho_\Lambda$.  For positive
  $\rho_\Lambda$, $\rho_{\rm vir}$ asymptotes to $16\rho_\Lambda$ at
  late times.  For negative $\rho_\Lambda$, the virial density
  formally goes to zero at $t=\pi t_\Lambda$, which the latest
  possible time the universe can crunch.  As we discuss in the text,
  our approximation (and, we shall assume, structure formation) fails
  when the virial density drops below the background matter density,
  which diverges near the crunch.}
\label{fig:VirialPlot}
\end{figure}

Note that according to this prescription, $\rho_{\rm vir}(t)$ has no
explicit dependence on the curvature of the background universe.
Changing $\rho_\Lambda$ results in a simple scaling behavior:
$\rho_{\rm vir}(t/t_{\Lambda})/|\rho_{\Lambda}|$ is independent of
$\rho_\Lambda$, where
\begin{equation}
t_\Lambda \equiv (3/8\pi\,G_{\rm N}\,|\rho_\Lambda|)^{1/2}~.
\end{equation}
In practice this means that one only has to compute $\rho_{\rm
  vir}(t)$ for two values of $\rho_\Lambda$ (one positive and one
negative).  In Fig.~\ref{fig:VirialPlot}, we show the
$\rho_\Lambda$-independent part of the virial density.

The virial density determines the gravitational timescale, $t_{\rm
  grav}$, which we define to be
\begin{equation}
t_{\rm grav} = \frac{1}{\sqrt{G_{\rm N} \rho_{\rm vir}}}~.
\end{equation}
This is the timescale for dynamical processes within the halo,
including star formation.

\subsection{Cooling and galaxy formation}

Halos typically virialize at a high temperature, $T_{\rm vir}$.
Before stars can form, the baryons need to cool to lower temperatures
and condense.  Cooling proceeds by bremsstrahlung, molecular or atomic
line cooling, and Compton cooling via scattering against the CMB.  

For bremsstrahlung and line cooling, the timescale for the baryons to
cool, $t_{\rm cool}$, is
\begin{equation}
t_{\rm cool} = \frac{3 T_{\rm vir}m_{\rm p}^2}{2 f_{\rm b} \mu X^2 \Lambda(T_{\rm vir}) \rho_{\rm vir}}
\end{equation}
with $\mu= 0.6\, m_{\rm p}$ the average molecular weight (assuming
full ionization), $X=.76$ the hydrogen mass fraction, and
$\Lambda(T_{\rm vir})$ the cooling function, which encodes the
efficiency of the bremsstrahlung and line cooling mechanisms as a
function of temperature.  For $T\lesssim10^4\,{\rm K}$, $\Lambda(T)$
is effectively zero and cooling does not occur.  When $T\gtrsim 10^7\,
{\rm K}$, bremsstrahlung cooling is the dominant mechanism and
$\Lambda(T)\propto T^{1/2}$.  For intermediate values of $T$,
$\Lambda(T)$ has a complicated behavior depending on the metallicity
of the gas~\cite{SutherlandDopita}.  For our purposes, however, it
suffices to approximate the cooling function as constant in this
range.  To summarize:
\begin{equation}
  \Lambda(T) = \begin{cases}
    0,   &\text{for $T<10^4\, {\rm K}$},\\
    \Lambda_0,  &\text{for $10^4\, {\rm K}<T<10^7\, {\rm K}$,}\\
    \Lambda_0\left(\frac{T}{10^7\, {\rm K}}\right)^{1/2},
    &\text{for $10^7\, {\rm K}<T$,}
\end{cases}
\end{equation}
with $\Lambda_0 = 10^{-23} {\rm erg}\,{\rm cm}^{3}\,{\rm s}^{-1}$.

We need to include the effects of Compton cooling separately.  In
general, Compton cooling depends on the temperature of the CMB,
$T_{\rm CMB}$, the temperature of the baryonic matter, and the
ionization fraction of the matter~\cite{White}.  But the dependence on
the gas temperature drops out when $T_{\rm vir}\gg T_{\rm CMB}$ (which
is always the case), and we will assume that the gas transitions from
neutral to completely ionized at $T_{\rm vir} = 10^4\,{\rm K}$.  With
these approximations, Compton cooling proceeds along the timescale
\begin{equation}
t_{\rm comp} = \frac{45 m_{\rm e}}{4\pi^2\sigma_{\rm T}  (T_{\rm CMB})^4},
\end{equation}
where $\sigma_{\rm T} = \frac{8\pi}{3}\left(\frac{e^2}{m_{\rm
      e}}\right)^2$ is the Thompson cross-section.  

The Compton timescale is independent of all of the properties of the
gas (provided $T_{\rm vir}>10^4{\rm K}$).  So this cooling mechanism
can still be efficient in regimes where other cooling mechanisms fail.
In particular, Compton cooling will be extremely effective in the
early universe while the CMB is still hot.  We define $\tau_{\rm
  comp}$ as the time when $t_{\rm comp}$ is equal to the age of the
universe.  $\tau_{\rm comp}$ is a reasonable convention for the time
beyond which Compton cooling is no longer efficient.  While we compute
$\tau_{\rm comp}$ numerically in practice, it is helpful to have an
analytic formula for a special case.  In a flat, $\rho_\Lambda=0$
universe we obtain
\begin{equation}
\tau_{\rm comp} =\left(\frac{4\pi^2\sigma_{\rm T}  (T_{\rm eq})^4}{45 m_{\rm e}}\left(\frac{t_{\rm eq}}{2-\sqrt{2}}\right)^{8/3}\right)^{3/5}\approx 0.763\left(\frac{T_{\rm eq}}{{\rm eV}}\right)^{-4/5}\,{\rm Gyr}\approx 0.9 \,{\rm Gyr}.
\end{equation}
Turning on strong curvature or $\rho_\Lambda$ will tend to lower $\tau_{\rm comp}$.

Before $\tau_{\rm comp}$, all structures with $T_{\rm vir}>10^4{\rm
  K}$ can cool efficiently and rapidly, so star formation will proceed
on the timescale $t_{\rm grav}$.  After $\tau_{\rm comp}$, some halos
will have $t_{\rm cool}<t_{\rm grav}$ and hence cool efficiently, and
some will have $t_{\rm cool}>t_{\rm grav}$ and be ``cooling-limited''.
We take it as a phenomenological fact that the cooling-limited halos
do not form stars. (One could go further and define a ``cooling
radius'' to take into account the radial dependence of the baryonic
density~\cite{White}, but this effect is negligible within our
approximations.)

In the absence of Compton cooling, then, we will consider only halos
for which $t_{\rm cool}<t_{\rm grav}$, which have masses in a finite range:
\begin{equation}
  M_{\rm  min}(t_{\rm vir})<M<M_{\rm max}(t_{\rm vir})~.
\end{equation}
The minimum halo mass at time $t_{\rm vir}$ corresponds to the minimum
temperature $T_{\rm vir}=10^4 {\rm K}$:
\begin{equation}
  M_{\rm min}(t_{\rm vir})=\left(\frac{5\times 10^4\,\mbox{K}}{G_{\rm N}\,\mu}\right)^{3/2}\left(\frac{3}{4\pi\rho_{\rm vir}}\right)^{1/2}~,
\end{equation}
where we have used the relation between virial temperature, density,
and mass for a uniform spherical halo derived from the virial theorem:
\begin{equation}
  T_{\rm vir} = \frac{\,G_{\rm N}\,\mu}{5}\left(\frac{4\pi}{3}\rho_{\rm vir}M^2\right)^{1/3}~.
\end{equation}
(A non-uniform density profile will result in a change to the
numerical factor on the RHS of this equation, but such a change makes
little difference to the result.)  

The maximum mass haloes at time $t_{\rm vir}$ satisfy $t_{\rm
  cool}=t_{\rm grav}$.  For $T<10^7 {\rm K}$, this mass is given by
\begin{equation}
  M_{\rm max}(t_{\rm vir}) =
  \left(\frac{6}{\pi}\right)^{1/2}
  \left(\frac{5\,f_{\rm b}X^2\Lambda_0}{3\,
      G_{\rm N}^{3/2}\,m_{\rm p}^2}\right)^{3/2}\rho_{\rm vir}^{1/4}~.
\end{equation}
Note that halos with mass $M_{\rm max}(t)$ have virial temperatures
less than $10^7 {\rm K}$ for $t\gtrsim0.24\,{\rm Gyr}$, which is well
before $\tau_{\rm comp}$.  This means that in our calculations we can
safely ignore the decreased cooling efficiency for extremely hot
halos.

\subsection{Star formation}

For halos satisfying the criterion $t_{\rm cool}<t_{\rm grav}$, the
hot baryonic material is allowed to cool.  Gravitational collapse
proceeds on the longer timescale $t_{\rm grav}$.  We will assume that
the subsequent formation of stars is also governed by this
timescale. Thus, we write the star formation rate for a single halo of
mass $M$ as
\begin{equation}\label{eq:SingleHaloSFR}
\frac{dM_\star^{\rm single}}{dt}(M, t_{\rm vir}) = A \frac{M_{\rm b}}{t_{\rm grav}(t_{\rm vir})} = A \frac{f_{\rm b}M}{t_{\rm grav}(t_{\rm vir})}~,
\end{equation}
where $M_{\rm b}$ is the total mass of baryons contained in the halo
and $f_{\rm b}$ is the baryon mas fraction.  We fix $f_{\rm b}$ to the
observed value $f_{\rm b} \approx 1/6$.\footnote{It would be
  interesting to consider variations of $f_{\rm b}$ in future work;
  see Ref.~\cite{Ben2008} for an environmental explanation of the
  observed value.}  

The order-one factor $A$ parametrizes the detailed physics of
gravitational collapse and star formation.  The free-fall time from
the virial radius to a point mass $M$ is $(3\pi/32)^{1/2}\,t_{\rm
  grav}$, and so we choose to set $A=(32/3\pi)^{1/2}$.  The final SFR
calculated for our universe is not very sensitive to this choice.
Notice that the single halo star formation rate depends only on the
mass and virialization time of the halo.

The next step in computing the complete SFR is to average over the
virialization times of the halos in existence at time $t$.  Define
$\frac{dM_\star^{\rm avg}}{dt}(M,t)$ to be the star formation rate of
all halos that have mass $M$ at time $t$, averaged over possible
virialization times $t_{\rm vir}$.  Now we need to consider the
possible virialization times.

It may happen that some halos in existence at time $t$ are so old that
they have already converted all of their baryonic mass into stars.
These halos should not be included in the calculation.  Furthermore,
feedback effects reduce the maximum amount of baryonic matter that can
ever go into stars to some fraction $f$ of the total.  ($f$ is a free
parameter of the model, and we shall fix its value by normalizing the
integrated SFR for our universe; see below.)  We should then only
consider halos with $t_{\rm vir}>t_{\rm min}$, where $t_{\rm min}$ is
a function of $t$ only and satisfies
\begin{equation}
\label{eq:tmin}
t-t_{\rm min} = \left(\frac{1}{f M_{\rm b}} \frac{dM_\star^{\rm single}}{dt}(M, t_{\rm min}) \right)^{-1} = \frac{f}{A} t_{\rm grav}(t_{\rm min})~.
\end{equation}
Though an explicit expression for $t_{\rm min}$ can be found in some
special cases, in practice we will always solve for it numerically.

We still need to account for the fact that some halos existing at $t$
formed at $t_{\rm vir}$ with masses outside the range $M_{\rm
  min}(t_{\rm vir})<M<M_{\rm max}(t_{\rm vir})$.  Since $M_{\rm min}$
is an increasing function of time while $M_{\rm max}$ is a decreasing
function of time, this condition on the mass range is equivalent to an
upper bound on $t_{\rm vir}$: $t_{\rm vir}<t_{\rm max}$, where $t_{\rm
  max}$ satisfies either
\begin{equation}
\label{eq:tmax1}
M = M_{\rm max}(t_{\rm max})
\end{equation}
or
\begin{equation}
\label{eq:tmax2}
M = M_{\rm min}(t_{\rm max})~,
\end{equation}
or else $t_{\rm max}=t$ if no solution exists.  Recall that the
condition $M<M_{\rm max}$ is a condition on cooling failure, and so is
only applicable for $t_{\rm vir}>\tau_{\rm comp}$.  Halos with $t_{\rm
  vir}<\tau_{\rm comp}$ have no upper mass limit.  Like $t_{\rm min}$,
$t_{\rm max}$ is found numerically.  Unlike $t_{\rm min}$, $t_{\rm
  max}$ is a function of both mass and time.

Now we can use the distribution on halo formation times deduced from
Eq.~\ref{eq:EPSFunction} to get
\begin{equation}\label{eq:AvgHaloSFR}
  \frac{dM^{\rm avg}_\star}{dt}(M,t)= 
  \int_{t_{\rm min}}^{t_{\rm max}} \;
  \left[ \frac{Af_{\rm b}M}{t_{\rm grav}(t_{\rm vir})}\,
    \frac{\partial P}{\partial t_{\rm vir}}(t_{\rm vir},M,t)\right]\,
  dt_{\rm vir}~.
\end{equation}
The final step is to multiply this average by the number of
halos of mass $M$ existing at time $t$ and add up the contributions
from all values of $M$.  Equivalently, we can multiply the normalized
average star formation rate, $\frac{1}{M}\frac{dM_\star^{\rm
    avg}}{dt}$, by the total amount of mass in halos of mass $M$ at time $t$ and
then add up the contributions from the different mass scales.  The
Press-Schechter formalism tells us that the fraction of the total mass
contained in halos of masses between $M$ and $M+dM$ is $\frac{\partial
  F}{\partial M}dM$, where $F(M,t)$ is the Press-Schechter mass
function defined in Eq.~\ref{eq:PSFraction}.  So restricting ourselves
to a unit of comoving volume with density $\rho_0$ (sometimes called
the reference density), we find that the SFR is given by
\begin{equation}
\dot{\rho}_\star(t)= \rho_0 \int  dM \int_{t_{\rm min}}^{t_{\rm max}} dt_{\rm vir}\, \frac{\partial F}{\partial M}(M,t)  \frac{Af_{\rm b}}{t_{\rm grav}(t_{\rm vir})} \frac{\partial P}{\partial t_{\rm vir}}(t_{\rm vir},M,t)~.
\end{equation}

\paragraph{Normalizing the SFR}

Our model contains the free parameter $f$ representing the maximum
fraction of the total baryonic mass of a halo which can go into stars.
We will impose a normalization condition on the SFR for our universe
to fix its value.  Nagamine {\em et al.}~\cite{Nagamine} normalize
their ``fossil" SFR so that
\begin{equation}
\int_0^{t_0} \dot{\rho}_{\star}\,dt = \frac{f_{\rm b} \rho_{\rm m}}{10},
\end{equation}
with $t_0=13.7\,{\rm Gyr}$ the age of the universe.  In other words,
the total amount of mass going into stars should be one tenth of the
total baryonic mass in the universe.  Due to the phenomenon of
recycling, this number is more than the actual amount of matter
presently found in stars.  We find that with $f=0.35$, our model
satisfies this normalization condition.

\paragraph{Early and late time cutoffs}

Even after $t=t_{\rm eq}$, baryonic matter remains coupled to photons
until $t=t_{\rm rec}\approx 3.7\times 10^5 \,{\rm yr}$.  So even
though dark matter halos can and do begin to collapse before
recombination, stars will obviously be unable to form.  After
$t=t_{\rm rec}$, the baryons are released and will fall into the
gravitational wells of the pre-formed dark matter halos.  Some time
$\Delta t$ (of order $t_{\rm rec}$) later, the baryonic matter will
have fallen into the dark matter halos and can form stars normally. In
our model, we account for this effect by placing further conditions on
$t_{\rm min}$. First, if $t<2\,t_{\rm rec}$ then the SFR is set to
zero. Then, provided $t>2\,t_{\rm rec}$, we compute $t_{\rm min}$
according to Eq.~\ref{eq:tmin}. If we find that $t_{\rm min} <
2\,t_{\rm rec}<t$, then the computation of $t_{\rm i}$ is flawed and
we set $t_{\rm min}=t_{\rm eq}$ to reflect the fact that dark matter
halos have been growing since equality. If $t_{\rm min}>2\,t_{\rm
  rec}$, then the computation is valid and we keep the result.

For $\rho_\Lambda<0$ there is an additional cutoff we must impose.
One can see from Fig.~\ref{fig:VirialPlot} that halo virial density in
such a universe continually decreases over time even as the background
matter density increases (following turnaround).  Eventually we reach
a point where halos virialize with densities less than that of the
background, which is a signal of the breakdown of the spherical
collapse model.  

The problem is that formally, the virial radius of the halo is larger
than the radius of the ``hole'' that was cut out of the background and
which is supposed to contain the ``overdensity''.  Thus, virialization
really leads to a collision with the background, and not to a star
forming isolated halo.  Physically, this can be interpreted as a
breakdown of the FRW approximation: sufficiently close to the crunch,
the collapsing universe can no longer be sensibly described as
homogeneous on average.

Therefore, it makes sense to demand that the halo density be somewhat
larger than the background density.  To pick a number, we assume that
under normal circumstances, a halo would settle into an isothermal
configuration with a density proportional to $1/r^2$.  Therefore the
density on the edge of the halo will only be one third of the average
density.  If we demand a reasonable separation of scales between the
density of the halo edge and that of the background matter, an
additional factor of about $3$ is not uncalled for.  So we will impose
the requirement that the average density of the halo be a factor of
ten higher than the background matter density.

We set the SFR to zero in any halos which do not satisfy this
requirement.  We implement this through a
modification of the computation of $t_{\rm max}$.  The true $t_{\rm
  max}$, computed at time $t$, is actually the minimum of the result above
(Eqs.~\ref{eq:tmax1}--\ref{eq:tmax2}) and the solution to
\begin{equation}
  \rho_{\rm vir}(t_{\rm max}) = 10\,\rho_{\rm m}(t)~.
\label{eq-latetimecut}
\end{equation}

\section{Results}
\label{sec-examples}

In this section, we apply our model to estimate the star formation
rate, and the integrated amount of star formation, in universes with
various amounts of curvature, vacuum energy, and density contrast.

\begin{figure}
\center
\subfigure{\includegraphics[width=3 in]{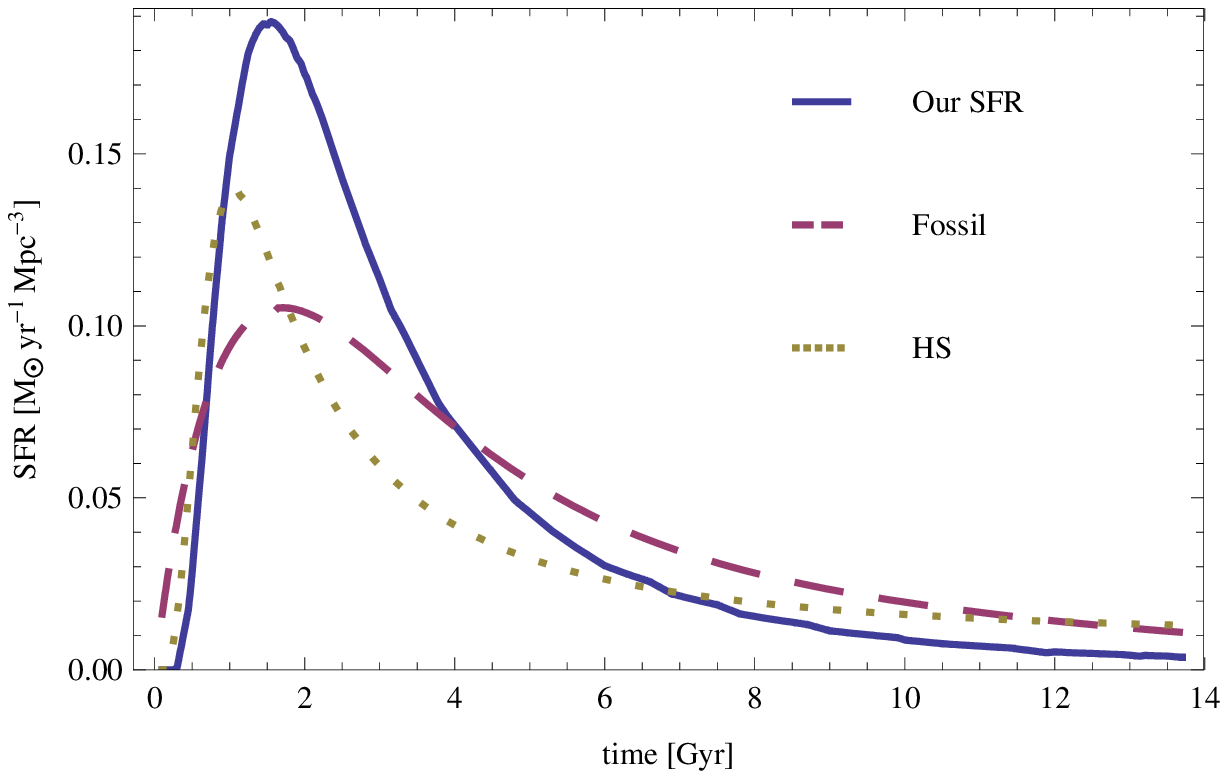}}
\subfigure{\includegraphics[width = 3 in]{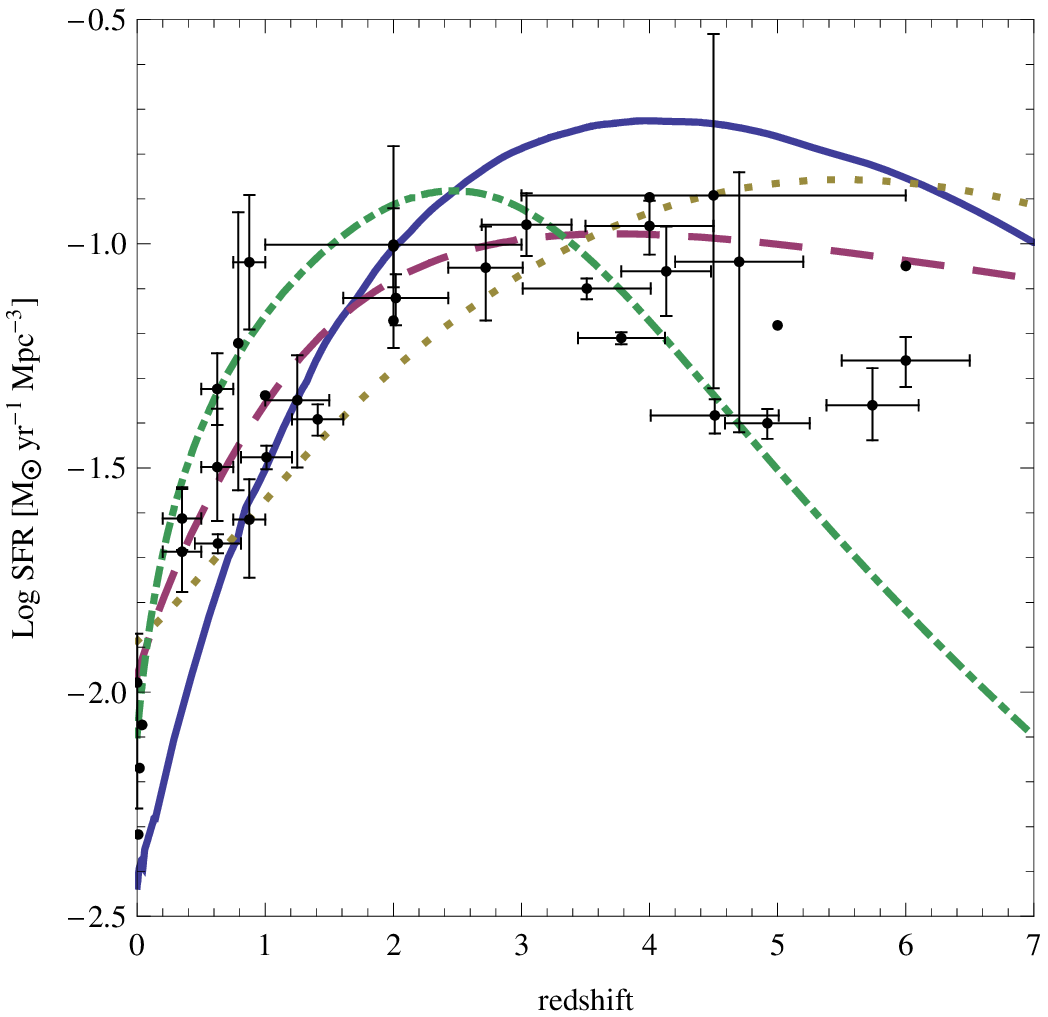}}
\caption{Comparison of our calculated SFR (solid) with the fossil
  model of Nagamine {\em et al.}~\protect\cite{Nagamine} (dashed) and
  the HS model~\protect\cite{HS} (dotted).  The right plot contains,
  in addition, the SFR of Hopkins and
  Beacom~\protect\cite{HopkinsBeacom} (dash-dotted).  The data points
  in the right plot are from astronomical observations and were
  compiled, along with references to their original sources,
  in~\protect\cite{Nagamine}}
\label{fig:SFRcomparison}
\end{figure}

\subsection{Our universe}

We begin by considering our own universe.
Figure~\ref{fig:SFRcomparison} shows the SFR computed by our model as
a function of time (left) and of redshift (right).  At intermediate
redshifts, our model is in good agreement with the data.  At other
redshifts, our model appears to be slightly off, but not by a large
factor, considering its crudeness.  For comparison, we show the SFR
predicted by the models of HS~\cite{HS}, Nagamine {\em
  et~al.}~\cite{Nagamine} and Hopkins and Beacom~\cite{HopkinsBeacom},
which contain a larger number of fitting parameters.

The shape of the SFR can be understood as a result of the interplay
between competing forces.  First, there will be no star formation
until structure forms with $T_{\rm vir}>10^4 {\rm K}$.  Once a
significant amount of structure begins to form at this critical
temperature, however, the SFR rises rapidly and soon hits its peak.

We can estimate the peak time, $t_{\rm peak}$, of the SFR by asking
when the typical mass scale that virializes is equal to $M_{\rm min}$:
\begin{equation}\label{eq:tpeak}
\frac{\delta_{\rm c}}{\sqrt{2} \sigma(M_{\rm min}(t_{\rm peak}),t_{\rm peak})}=1~.
\end{equation}
(There are ambiguities at the level of factors of order one; the
factor $\sqrt{2}$ was chosen to improve agreement with the true peak
time of the SFR.)  For the parameters of our universe, this
calculation yields $t_{\rm peak} = 1.7\,\mbox{Gyr}$, close to the peak
predicted by our model at about $1.6\,\mbox{Gyr}$.

After $t\approx t_{\rm peak}$, the SFR falls at least as fast as $1/t$
because $t_{\rm grav}\propto t$.  Once the typical virialized mass
exceeds the cooling limit, $M_{\rm max}$, the SFR falls faster still.
These considerations generalize to universes besides our own after
accounting for the peculiar effects of each of the cosmological
parameters.  In particular, as explained below, we should be able to
accurately estimate the peak time by this method for any universe
where the effects of curvature or the cosmological constant are
relatively weak.

\begin{figure}[h]
\center
\subfigure{
\includegraphics[width=3.5 in]{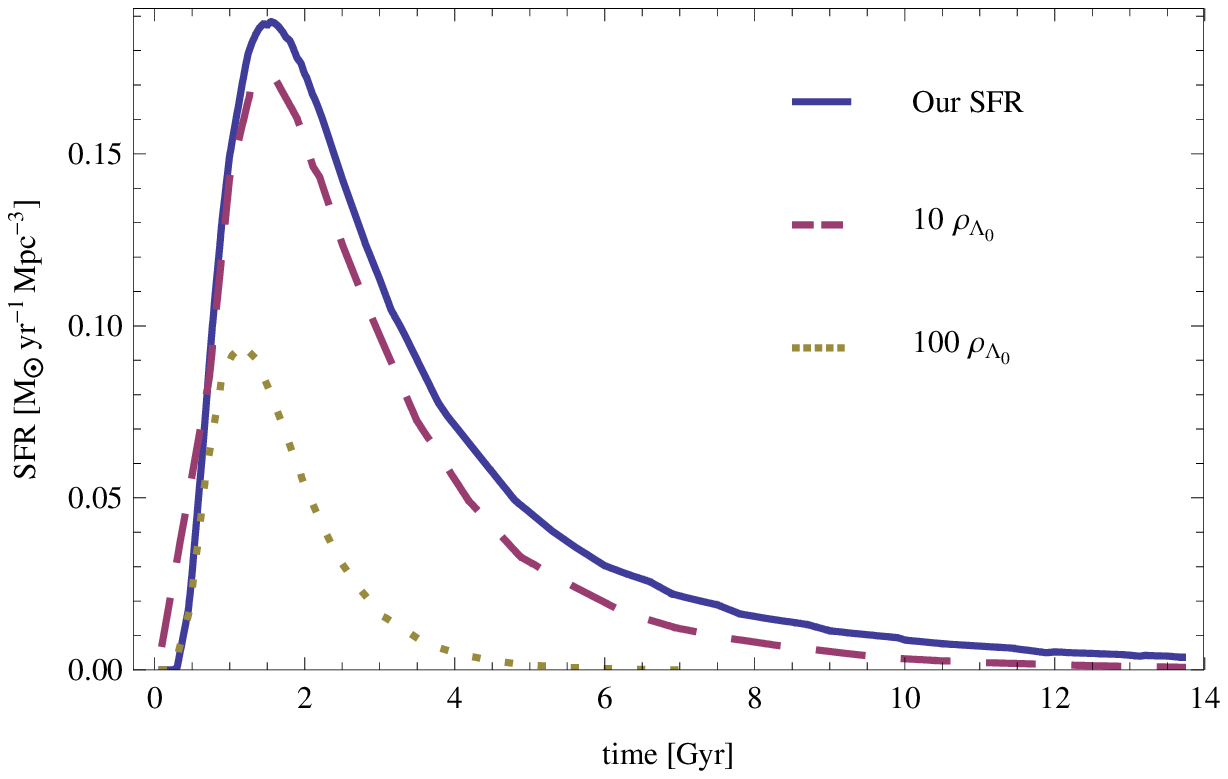}
}
\subfigure{
\includegraphics[width=3.5 in]{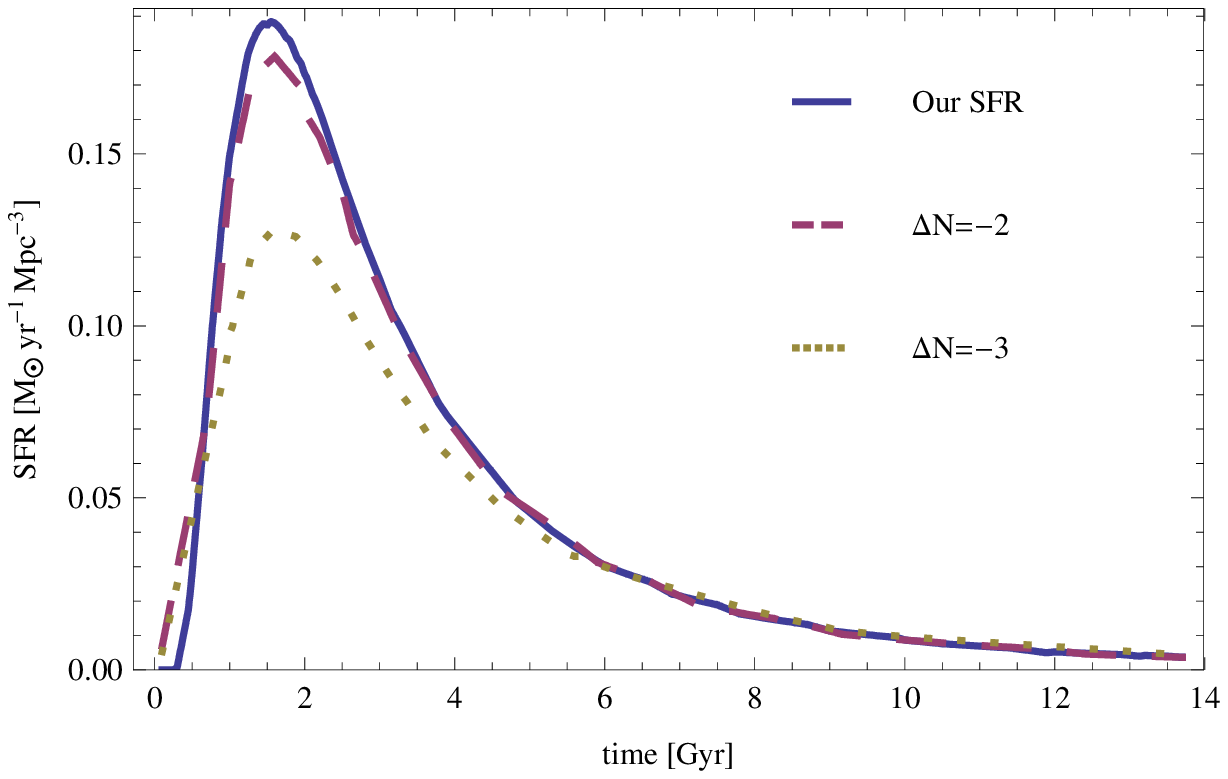}
}
\caption{The SFR for universes with greater cosmological constant
  (top) and more curvature (bottom) than ours.  All other parameters
  remain fixed to the observed values, with curvature assumed to be
  open and just below the observational bound.}
\label{fig:LambdaNPlots}
\end{figure}

\begin{figure}[h]
\center
\subfigure[$\Delta N$ with $\rho_\Lambda=0$]{
\includegraphics[width=2.5 in]{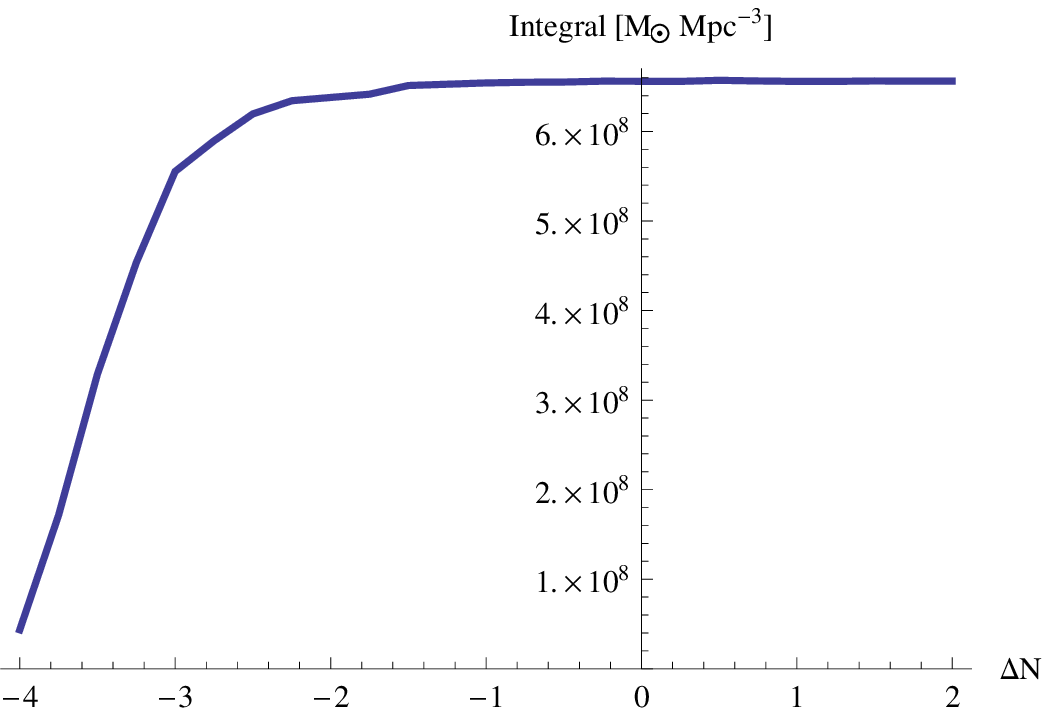}\label{fig:NIntegratedPlot}
}
\subfigure[$\rho_\Lambda > 0$]{
\includegraphics[width=2.7 in]{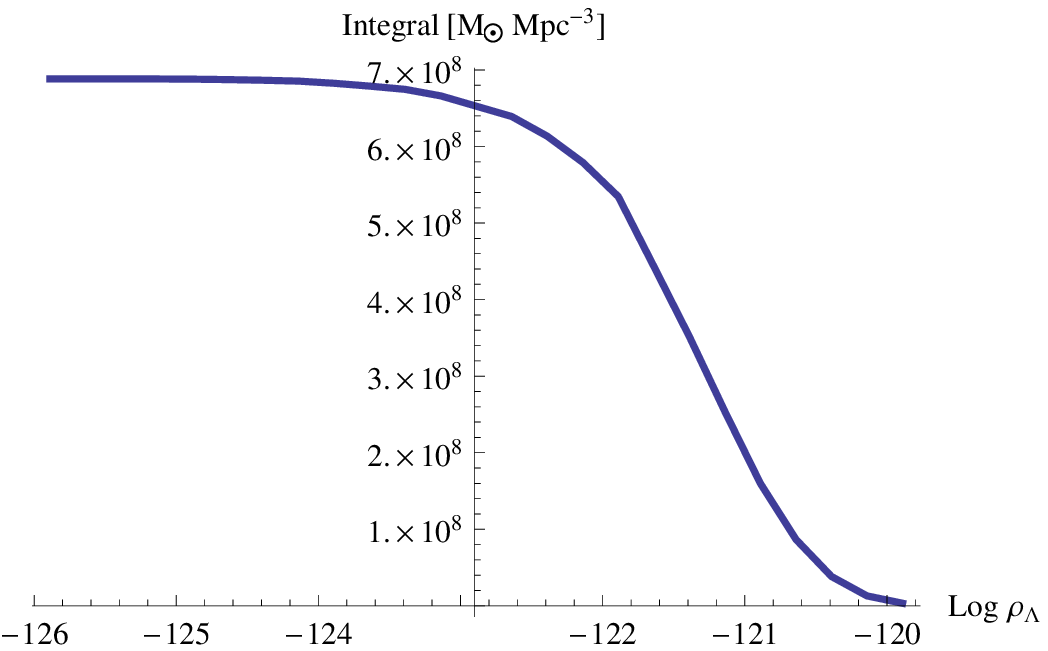}\label{fig:posLambdaIntegrated}
}
\subfigure[$\rho_\Lambda<0$]{
\includegraphics[width=2.5 in]{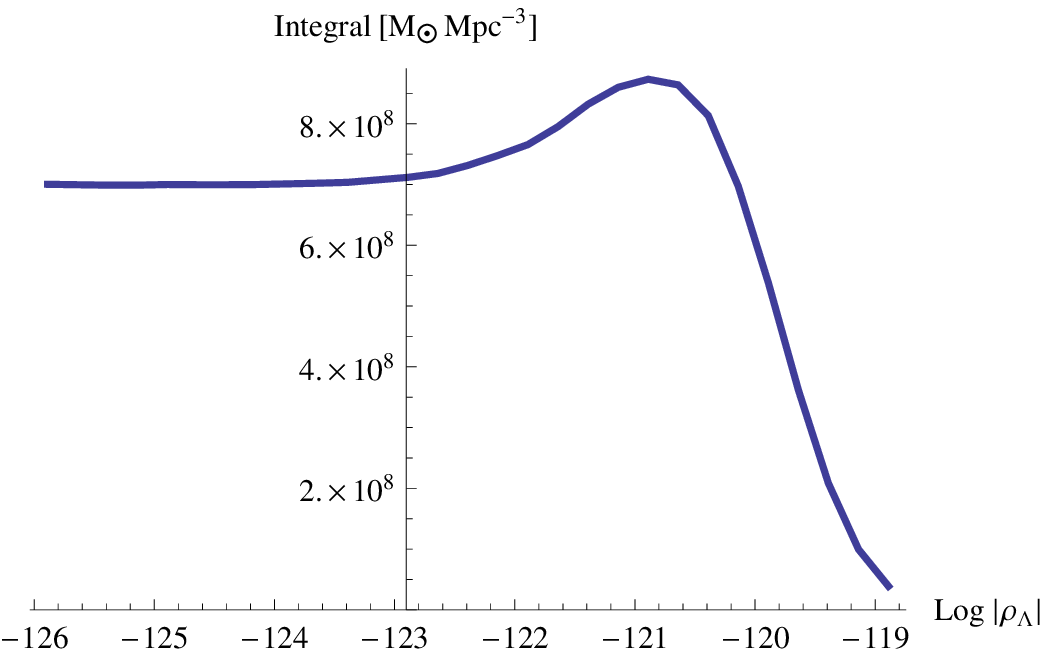}
}
\subfigure[All values of $\rho_\Lambda$]{
\includegraphics[width=2.5 in]{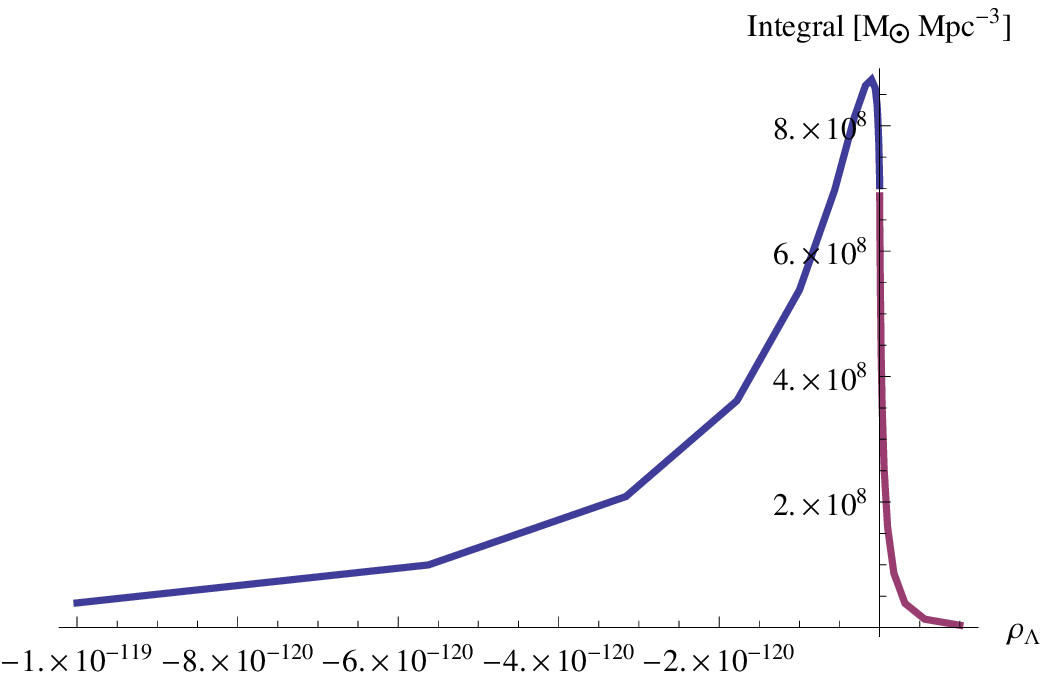}
}

\caption{The total stellar mass produced per comoving volume, as a
  function of open spatial curvature and the cosmological constant.
  (a) Curvature: Integrated SFR as a function of the number of efolds
  of inflation, relative to the observational bound on open spatial
  curvature.  For $\Delta N\gtrsim -2$, star formation is unaffected.
  More curvature (less inflation) suppresses star formation, as
  expected.  (The small irregularities in the asymptotic regime are
  numerical artifacts.)  (b) Positive cosmological constant:
  $\rho_\Lambda$ is given in Planck units; the observed value,
  $\rho_{\Lambda_0}\approx 1.25 \times10^{-123}$, is indicated by the
  position of the vertical axis.  Decreasing $\rho_\Lambda$ has no
  effect on star formation.  Larger values interfere with structure
  formation, and star formation becomes suppressed.  (c) Negative
  cosmological constant: Enhanced structure formation leads to more
  star formation for negative vacuum energy compared to positive
  vacuum energy of equal magnitude.  (d) Positive and negative values
  of the cosmological constant shown on a linear plot.}
\label{fig:IntegratedPlots}  
\end{figure}

\subsection{Varying single parameters}

Next, we present several examples of star formation rates computed
from our model for universes that differ from ours through a single
parameter.

\paragraph{Curvature and positive cosmological constant}

In Fig.~\ref{fig:LambdaNPlots}, we show the SFR for our universe
together with the SFRs computed for larger values of the cosmological
constant (top) and larger amounts of curvature (bottom).  One can see
that increasing either of these two quantities results in a lowered
star formation rate.\footnote{This differs from the conclusions
  obtained by Cline {\em et al.}~\cite{Cline} from an extrapolation of
  the Hernquist-Springel analytical model~\cite{HS}.  This model
  approximates all haloes in existence at time $t$ as freshly formed.
  Moreover, it does not take into account the decreased relative
  overdensity of virialized halos in a vacuum-dominated cosmology.
  These approximations are good at positive redshift in our universe,
  but they do not remain valid after structure formation ceases.}
Essentially this is because structure formation as a whole is
inhibited in these universes.  Star formation is obstructed by the
cosmological constant starting at the time $t\sim t_\Lambda\equiv
(3/\Lambda)^{1/2}$.  Open spatial curvature suppresses star formation
after the time of order $t_{\rm c}=0.533\frac{8\pi}{3}\rho_{\rm
  eq}a(t_{\rm eq})^3$, when $\Omega=0.5$ in an open universe with
$\rho_{\Lambda}=0$.  Not shown are SFRs from universes which have a
smaller cosmological constant than our universe, nor universes that
are more spatially flat than the observational bound ($\Delta N>0$).
The SFRs for those choices of parameters are indistinguishable from
the SFR for our universe.

In Fig.~\ref{fig:NIntegratedPlot}, the integrated star formation rate
is shown as a function of $\Delta N$.  It is apparent that extra
flatness beyond the current experimental bound does not change star
formation.  Indeed, the curve remains flat down to about $\Delta
N\approx -2$, showing that the universe could have been quite a bit
more curved without affecting star formation.  Integrated star
formation is suppressed by a factor 2 for $\Delta N=-3.5$ and by a
factor 10 for $\Delta N=-4$.

These numbers differ somewhat from the catastrophic boundary used in
Ref.~\cite{FreivogelKleban}, $\Delta N=-2.5$.  This is because the
observational upper bound on $\Omega_k$ has tightened from the value
$20\times 10^{-3}$ used in Ref.~\cite{FreivogelKleban} to $1.1\times
10^{-3}$.  The observationally allowed region corresponds to $\Delta
N\geq 0$.  Future probes of curvature are limited by cosmic variance
to a sensitivity of $\Omega_k\gtrsim 10^{-4}$ or worse, corresponding
to $\Delta N\lesssim 1$ at most.  The window for discovery of open
curvature has narrowed, but it has not closed.

The string landscape predicts a non-negligible probability for a
future detection of curvature in this window.  (This prediction is
contingent on cosmological measures which do not reward volume
expansion factors during slow-roll inflation, such as the causal
diamond measure~\cite{Bou06} or the scale factor
measure~\cite{DeSimoneetal}; such measures are favored for a number of
reasons.)  Depending on assumptions about the underlying probability
distribution for $\Delta N$ in the landscape (and on the details of
the measure), this probability may be less than 1\% or as much as
10\%~\cite{Futurepaper}.  However, the fact that curvature is already
known to be much less than the anthropic bound makes it implausible to
posit a strong landscape preference for small $\Delta N$, making a
future detection rather unlikely.

In Fig.~\ref{fig:posLambdaIntegrated}, the integrated star formation
rate is shown as a function of (positive) $\rho_\Lambda$. The observed
value of $\rho_\Lambda$ is right on the edge of the downward slope,
where vacuum energy is beginning to affect star formation. Integrated
star formation is suppressed by a factor 2 for $\rho_\Lambda=36
\rho_{\Lambda_0}$ and by a factor 10 for $\rho_\Lambda= 225
\rho_{\Lambda_0}$. To obtain a probability distribution for $\rho_\Lambda$
from this result, one needs to combine it with a cosmological
measure~\cite{Futurepaper}.

\begin{figure}[h]
\center
\subfigure[$\abs{\rho_\Lambda} = \rho_{\Lambda_0}$]{
\includegraphics[width=2.5 in]{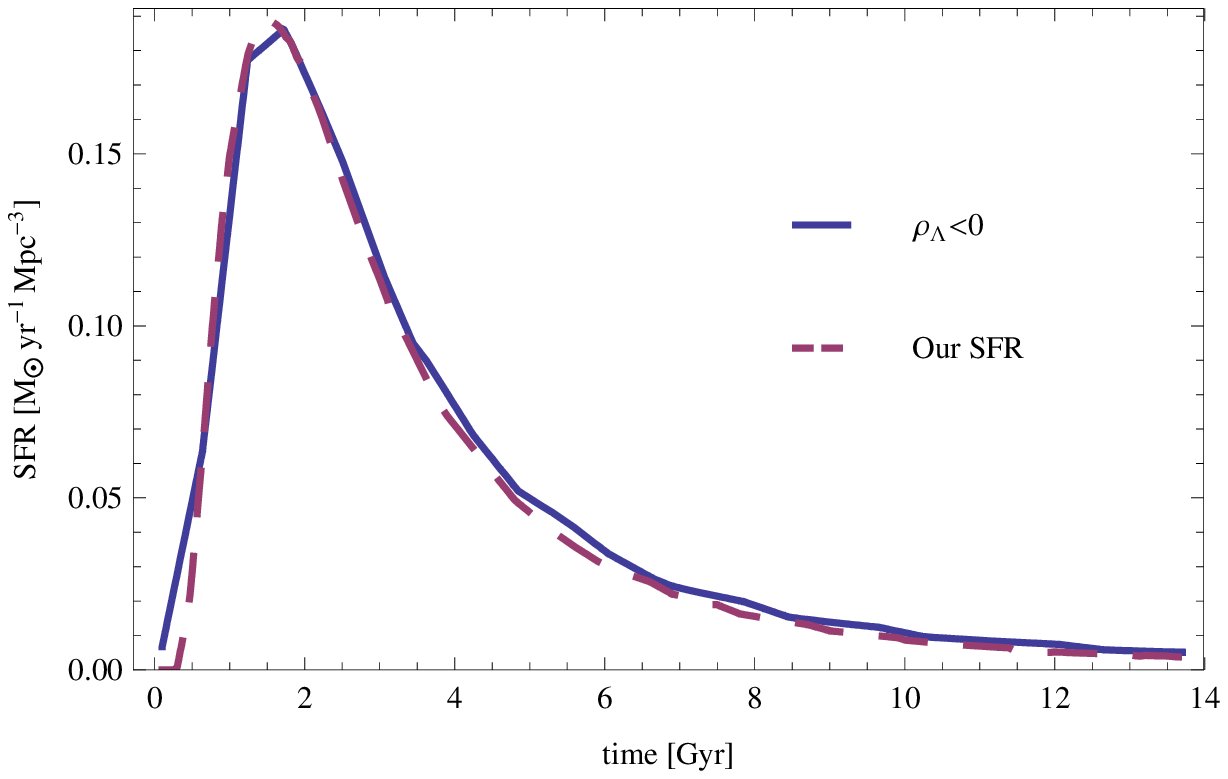}
}
\subfigure[$\abs{\rho_\Lambda} = 100 \, \rho_{\Lambda_0}$]{
\includegraphics[width=2.5 in]{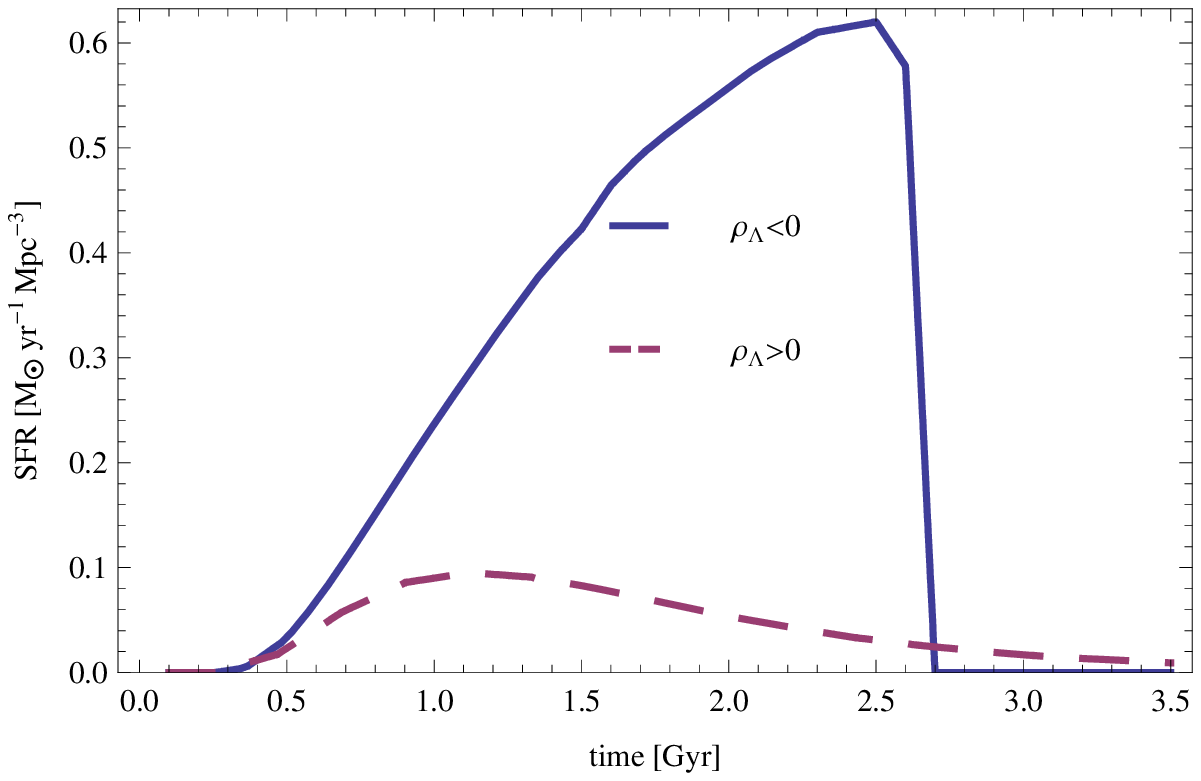}\label{fig:LargeHump}
}
\subfigure[$\Delta N = -4$, $\abs{\rho_\Lambda}=\rho_{\Lambda_0}$]{
\includegraphics[width=2.5 in]{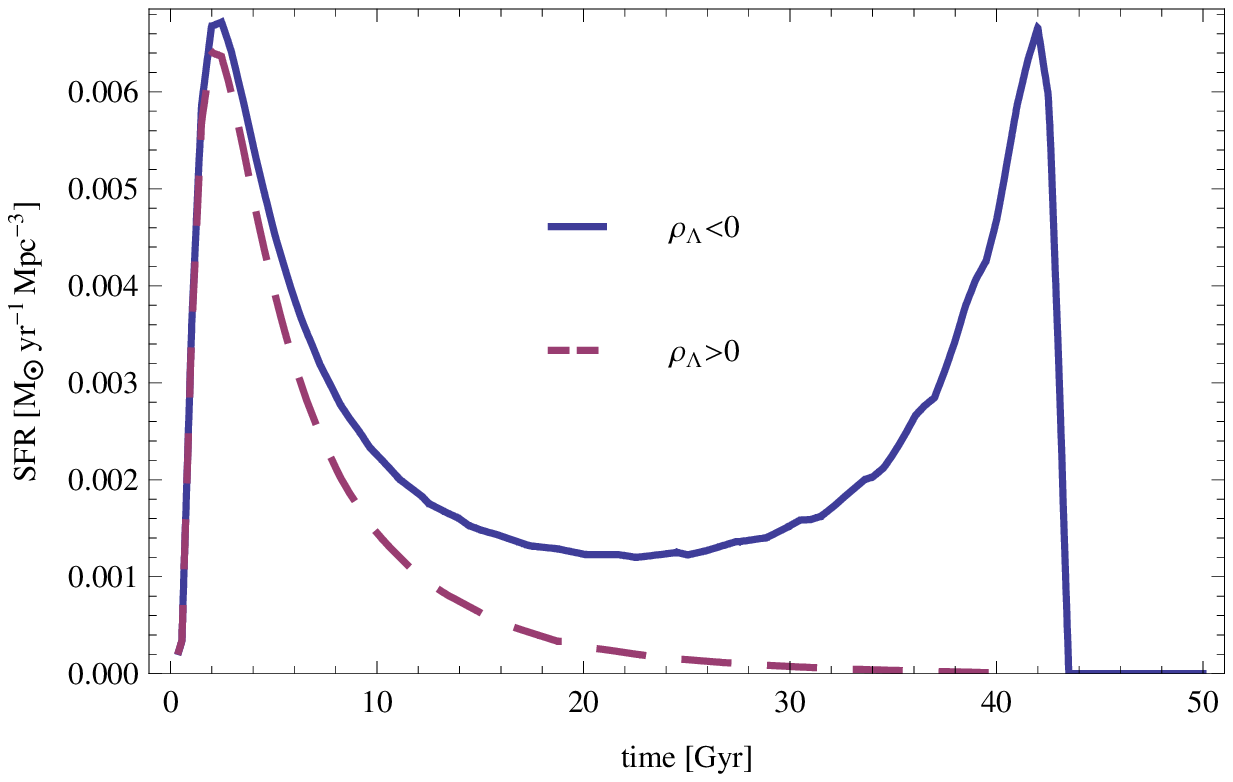}\label{fig:SecondPeak}
}
\caption{The SFR with negative cosmological constant.  In each figure
  we consider two different universes whose parameters are identical
  except for the sign of $\rho_\Lambda$.  (a) Changing the sign of the
  observed cosmological constant would have made little difference to
  the SFR.  However, in universes with a higher magnitude of
  cosmological constant (b), or more highly curved universes (c), a
  negative cosmological constant allows more star formation than a
  positive one of equal magnitude.  The apparent symmetry in (c) is
  accidental.}
\label{fig:negLambdaPlots}
\end{figure}

\begin{figure}[h]
\center
\subfigure{
\includegraphics[width=3.5 in]{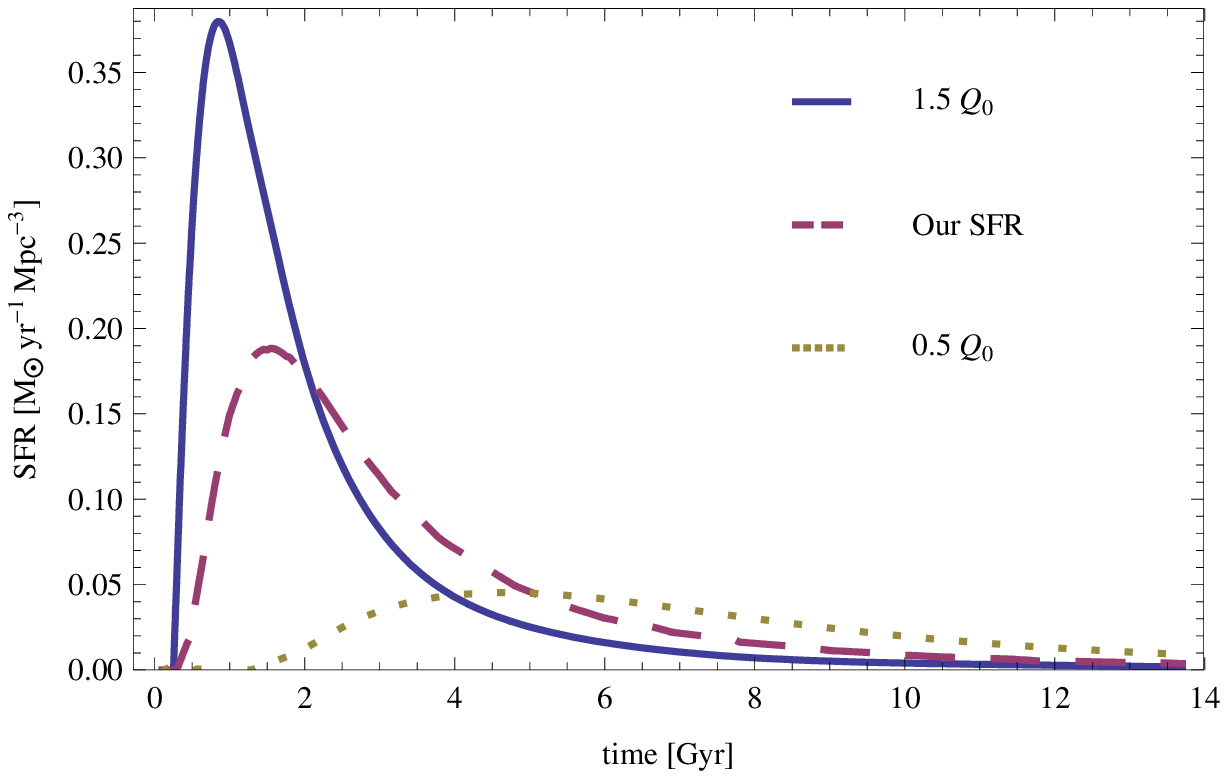}
}
\subfigure{
\includegraphics[width=3.5 in]{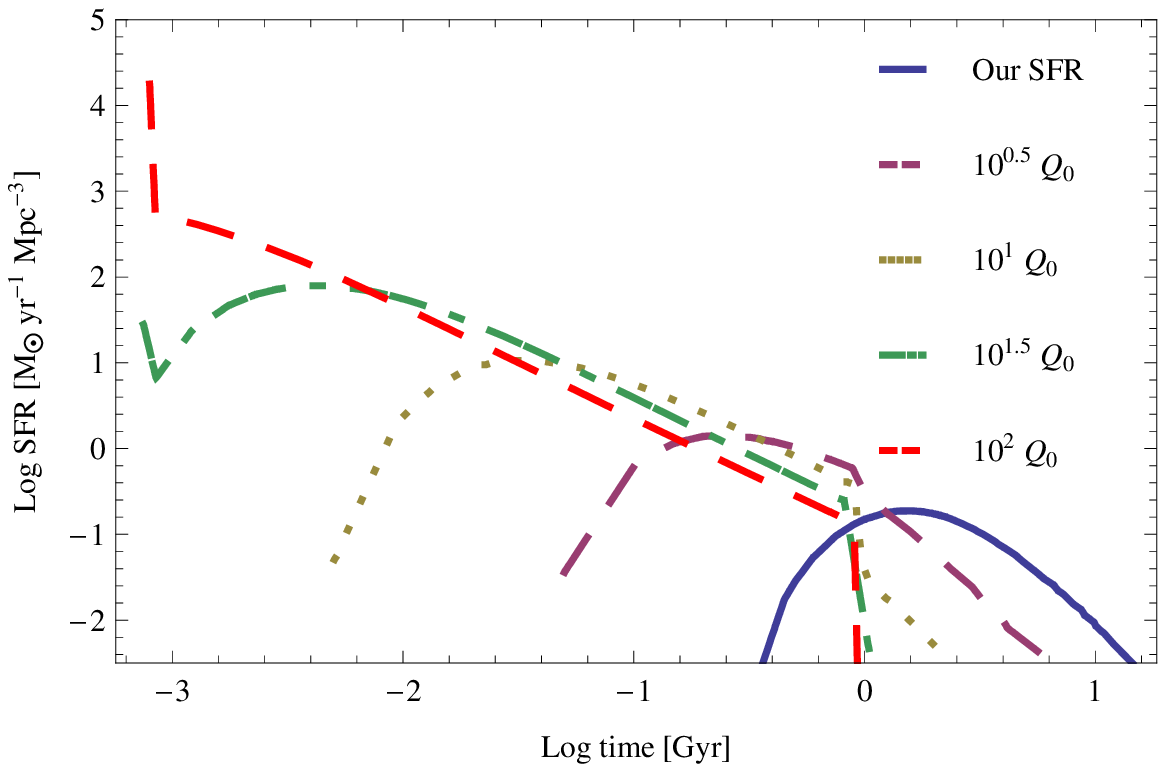}
}
\caption{The SFR for different values of the primordial density
  contrast, $Q$.  The upper plot shows the great sensitivity of the
  SFR to relatively small changes in $Q$.  Compton cooling plays no
  role here.  The lower plot shows a larger range of values.  The
  universal slope of $-1$ in the log-log plot corresponds to the $1/t$
  drop-off in the Compton cooling regime.  When Compton cooling
  becomes ineffective at around $1$ Gyr, most of the SFRs experience a
  sudden drop. The spike at early times for large values of $Q$ arises
  from the sudden influx of baryons after recombination.}
\label{fig:SFRQPlots}
\end{figure}
\begin{figure}[h]
\center
\includegraphics[width=3.5 in]{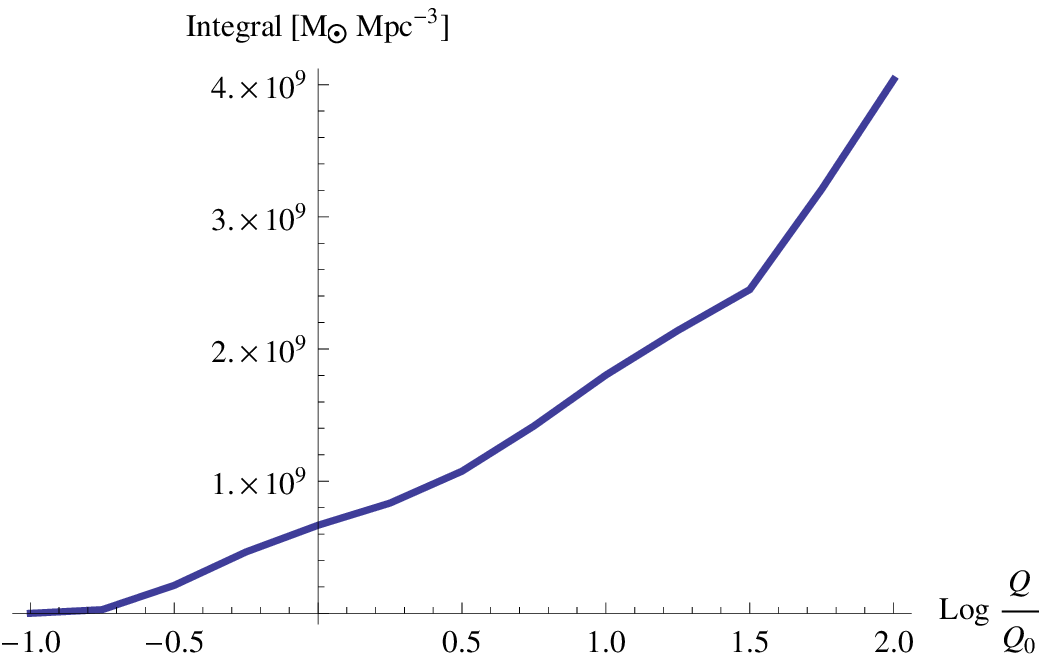}
\caption{In our model, total star formation increases with the density
  contrast.  This increase continues until we can no longer trust our
  SFR due to the dominance of large-scale structure at recombination.
  As discussed in the text, a more refined model may find star
  formation inhibited in parts of the range shown here.}
\label{fig:QIntegratedPlot}
\end{figure}


\paragraph{Negative cosmological constant}

In Fig.~\ref{fig:negLambdaPlots} we see the SFR in some universes with
negative cosmological constant. It is instructive to compare universes
that differ only through the sign of the cosmological constant. The
universe with negative vacuum energy will eventually reach a Big
Crunch, where star formation ends along with everything else.  Yet, it
will generally have a greater star formation rate than its partner
with positive vacuum energy. This is because in the positive case,
structure formation is severely hindered as soon as vacuum energy
comes to dominate, at $t\approx t_\Lambda$, due to the exponential
growth of the scale factor. In the negative case, this timescale
roughly corresponds to the turnaround. Structure formation not only
continues during this phase and during the early collapse phase, but
is enhanced by the condensation of the background.  Star formation is
eventually cut off by our requirement that the virial density exceed
the background density by a sufficient factor,
Eq.~(\ref{eq-latetimecut}).  This is the origin of the precipitous
decline in Fig.~\ref{fig:negLambdaPlots}(b) and (c).

The integrated star formation rate as a function of negative values of
the cosmological constant is shown in
Fig.~\ref{fig:IntegratedPlots}. As in the positive case, sufficiently
small values of $|\rho_\Lambda|$ do not affect star formation. A
universe with large negative $\rho_\Lambda$ will crunch before stars
can form.  For intermediate magnitudes of $\rho_\Lambda$, there is
more star formation in negative than in positive cosmological constant
universes.  The amount of this excess is quite sensitive to our cutoff
prescription near the crunch, Eq.~(\ref{eq-latetimecut}).  With a more
lenient cutoff, there would be even more star formation in the final
collapse phase.  A stronger cutoff, such as the condition that
$\rho_{\rm vir}>100\,\rho_{\rm m}$, would eliminate the excess
entirely.  Clearly, a more detailed study of structure and star
formation in collapsing universes would be desirable.

\paragraph{Density contrast}

In Fig.~\ref{fig:SFRQPlots} we see the effect of increasing or
decreasing the amplitude of primordial density perturbations, $Q$.
Note that even relatively small changes in $Q$ have a drastic impact
on the SFR. Increasing $Q$ has the effect of accelerating structure
formation, thus resulting in more halos forming earlier. Earlier
formation times mean higher densities, which in turn means higher star
formation rates.

Of course, higher $Q$ also leads to larger halo masses. One might
expect that cooling becomes an issue in these high-mass halos, and
indeed for high values of $Q$ the SFR drops to zero once Compton
cooling fails, at $t=\tau_{\rm comp}$. However, for $t<\tau_{\rm
  comp}$, cooling does not limit star formation, and so in high $Q$
universes there tends to be a very large amount of star formation at
early times. For the highest values of $Q$ that we consider, there is
a spike in the star formation rate at $t=2\,t_{\rm rec}$ coming from
the sudden infall of baryons to the dark matter halos. The spike is
short-lived and disappears once that initial supply of gas is used up.

In Fig.~\ref{fig:QIntegratedPlot} we show the integral of the SFR as a
function of $Q$. The increase in total star formation is roughly
logarithmic, and in our model the increase keeps going until most
structure would form right at recombination, which signals a major
regime change and a likely breakdown of our model.

A universe with high $Q$ is a very different place from the one we
live in: the densities and masses of galaxies are both extremely
high. It is possible that qualitatively new phenomena, which we are
not modelling, suppress star formation in this regime. For example, no
detailed study of cooling flows, fragmentation, and star formation in
the Compton cooling regime has been performed. We have also neglected
the effects of enhanced black hole formation.

\begin{figure}[h]
\center
\subfigure[]{
\includegraphics[width=2.5 in]{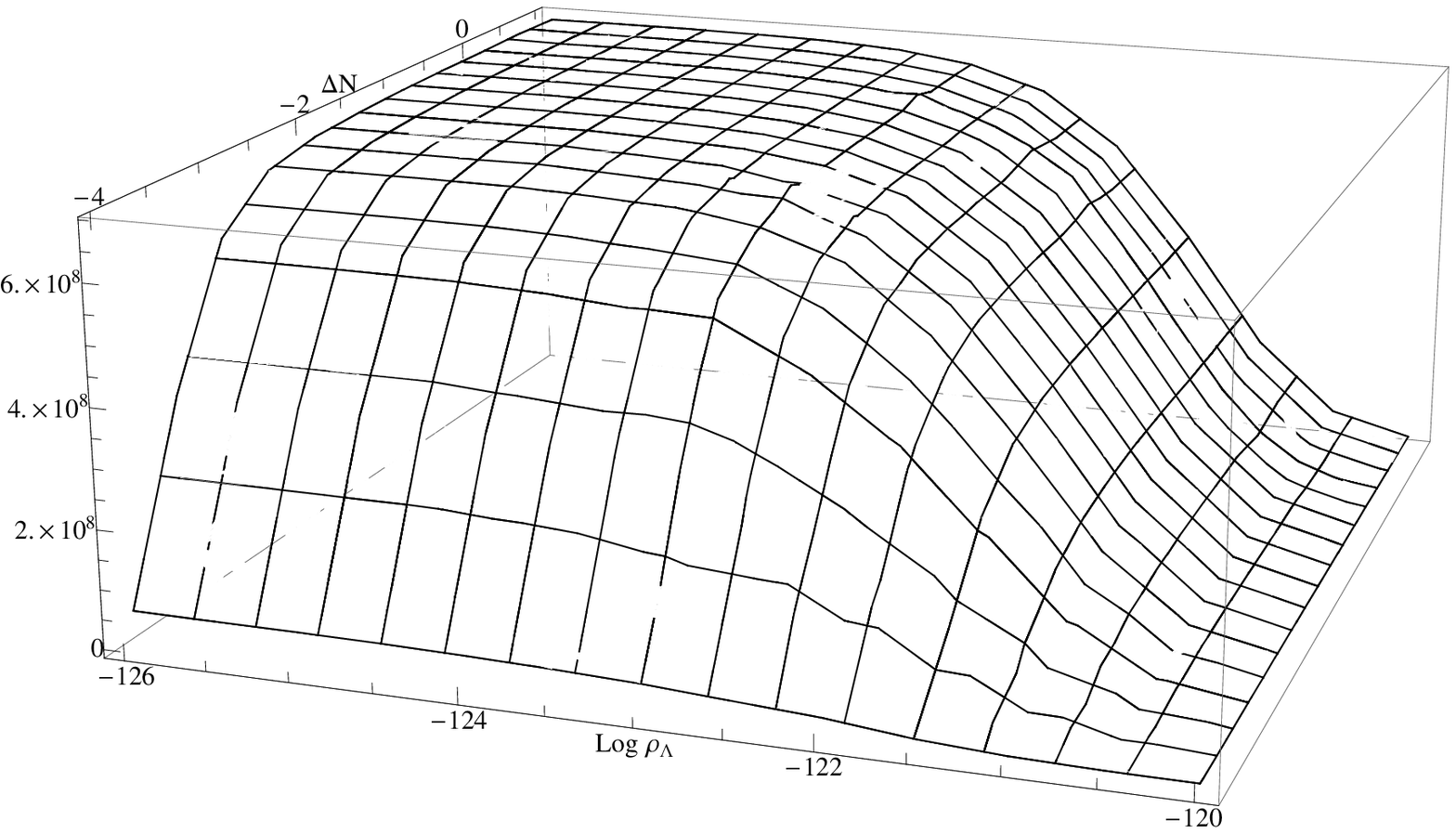}
}
\subfigure[]{
\includegraphics[width=2.5 in]{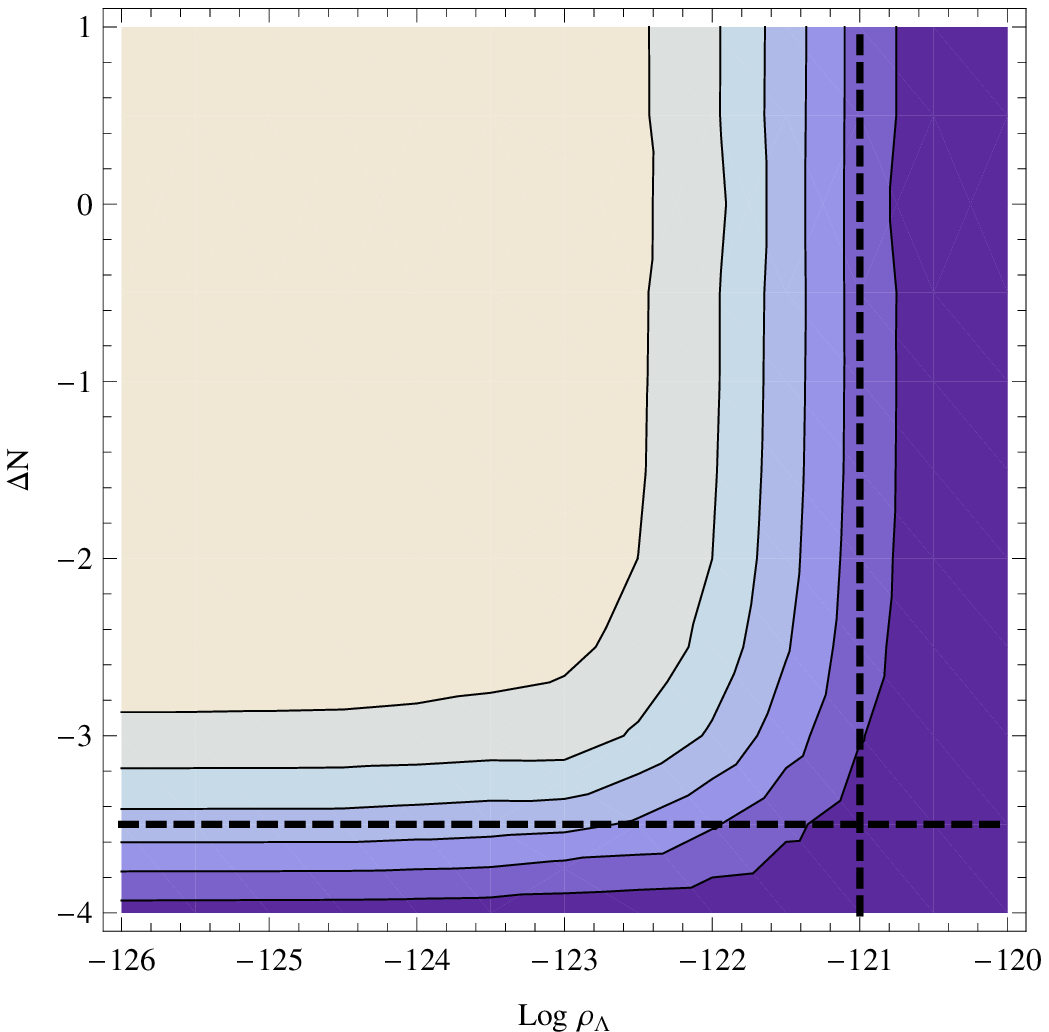}\label{fig:LambdaNContourPlot}
}
\caption{The integrated SFR as a function of curvature ($\Delta N$) and positive $\rho_\Lambda$, shown in a 3D plot (a) and in a contour plot (b).  (The strength of density perturbations, $Q$, is fixed to the observed value.)  Analytically one expects star formation to be suppressed if $t_{\rm peak}>t_{\rm c}$ or $t_{\rm peak}>t_\Lambda$, where $t_{\rm peak}$ is the peak star formation time in the absence of vacuum energy and curvature.  These catastrophic boundaries, $\log_{10} \rho_\Lambda = -121$ and $\Delta N = -3.5$, are shown in (b).
}
\label{fig:LambdaNIntegratedPlots}
\end{figure}

\begin{figure}[h]
\center
\subfigure[]{
\includegraphics[width=2.5 in]{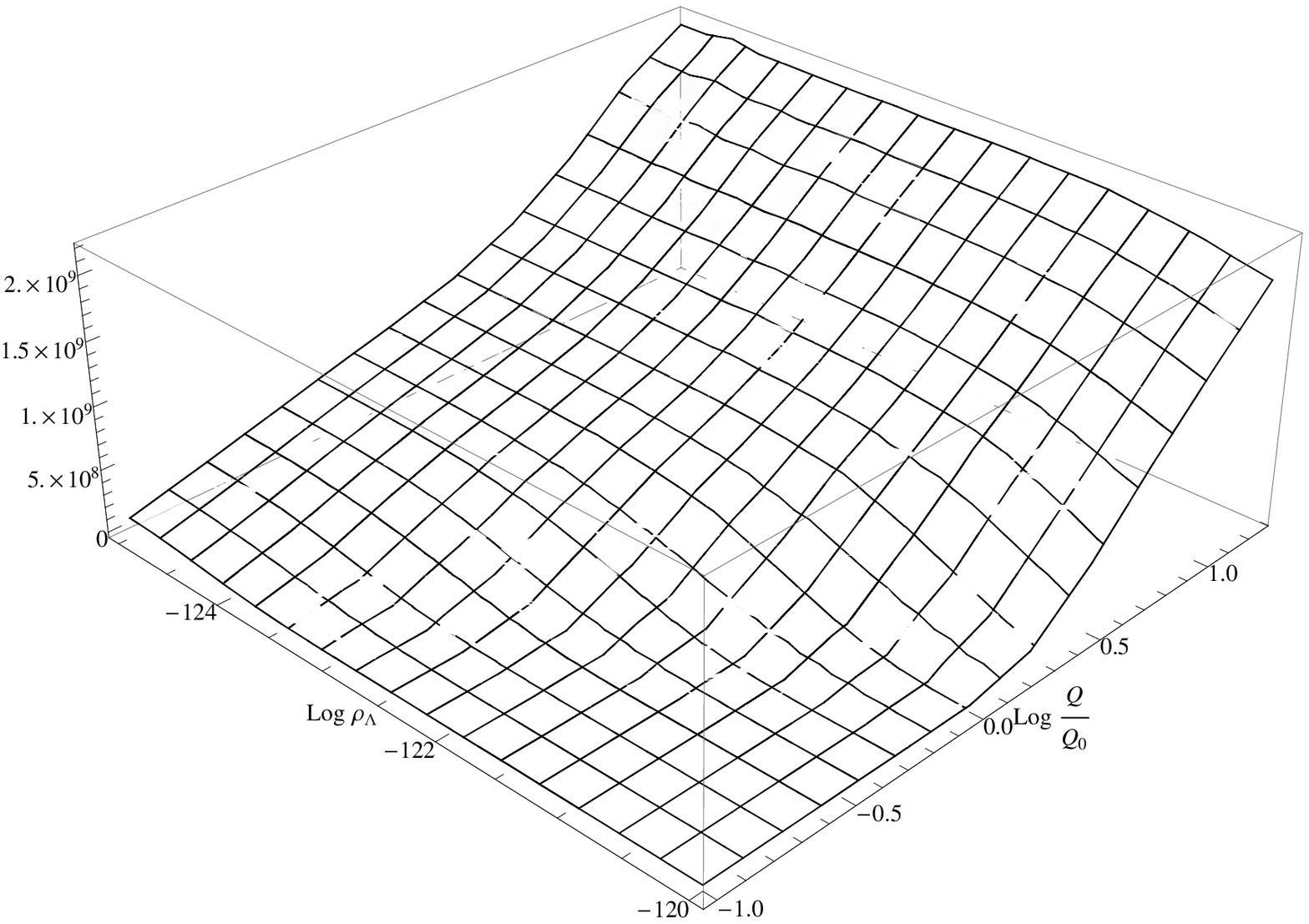}
}
\subfigure[]{
\includegraphics[width=2.5 in]{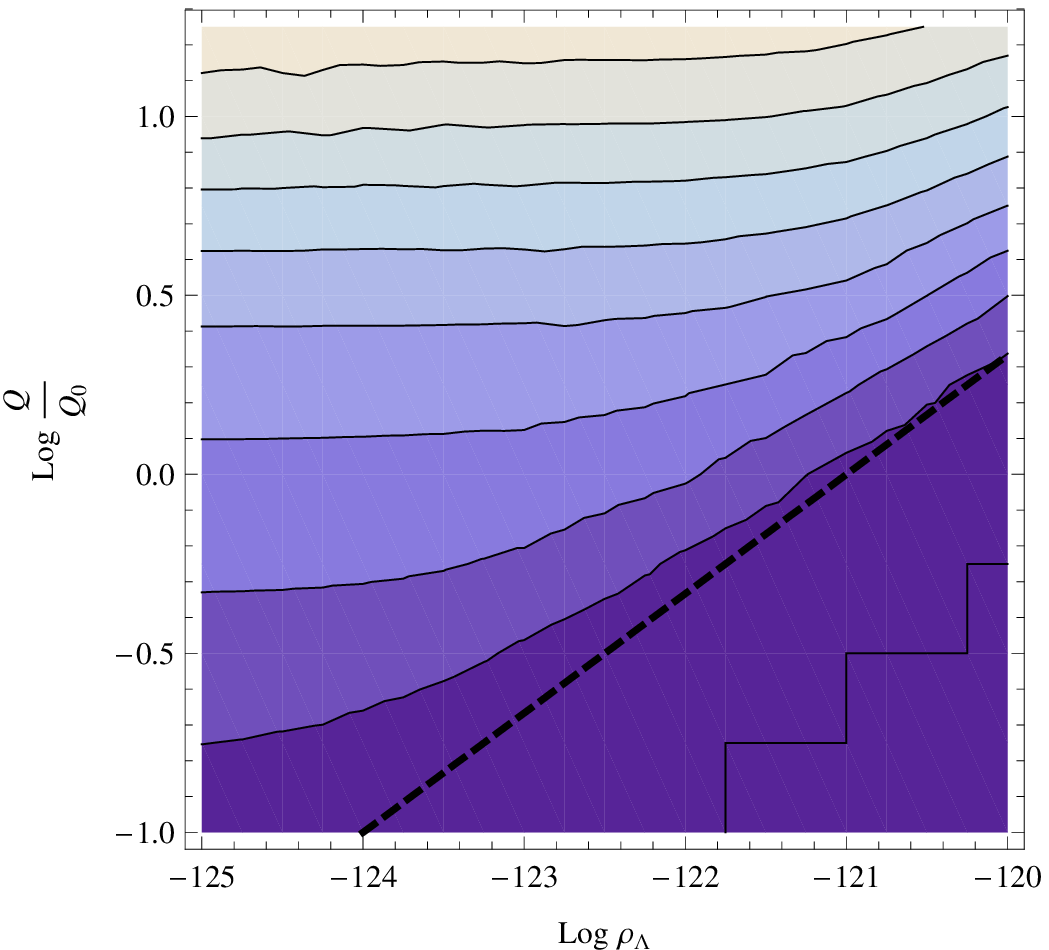}\label{fig:LambdaQContourPlot}
}
\caption{The integrated SFR as a function of positive $\rho_\Lambda>0$ and $Q$, at fixed spatial curvature, $\Delta N=0$~.  The contour plot (b) also shows the catastrophic boundary $t_{\rm peak}(Q)>t_\Lambda$ expected analytically.
}
\label{fig:LambdaQIntegratedPlots}
\end{figure}

\subsection{Varying multiple parameters}

We can also vary multiple parameters at once. For example, let us hold
only $Q$ fixed and vary both curvature and (positive) vacuum
energy.  The integrated star formation is is shown in
Fig.~\ref{fig:LambdaNIntegratedPlots}. If either curvature or
$\rho_\Lambda$ get large, then structure formation is suppressed and
star formation goes down. When the universe has both small
$\rho_\Lambda$ and small curvature, then structure formation is not
suppressed and the SFR is nearly identical to that of our universe.
Here ``large'' and ``small'' should be understood in terms of the
relation between $t_\Lambda$, $t_{\rm c}$, and $t_{\rm peak}$ computed
in a flat, $\rho_\Lambda=0$ universe (an unsuppressed universe). In
the case we are considering, namely $Q=Q_0$, the unsuppressed peak
time is $t_{\rm peak}=1.7\,\mbox{Gyr}$. The conditions $t_{\rm
  peak}=t_\Lambda$ and $t_{\rm peak} = t_{\rm c}$ translate into
$\rho_\Lambda \approx 100\,\rho_{\Lambda_0}$ and $\Delta N \approx
-3.5$~. These lines are marked in Fig.~\ref{fig:LambdaNContourPlot},
and one sees that this is a good approximation to the boundary of the
star-forming region.  (We will see below that this approximation does not continue to hold if both curvature and $Q$ are allowed to vary.)

\begin{figure}[h]
\center
\subfigure[]{
\includegraphics[width=2.5 in]{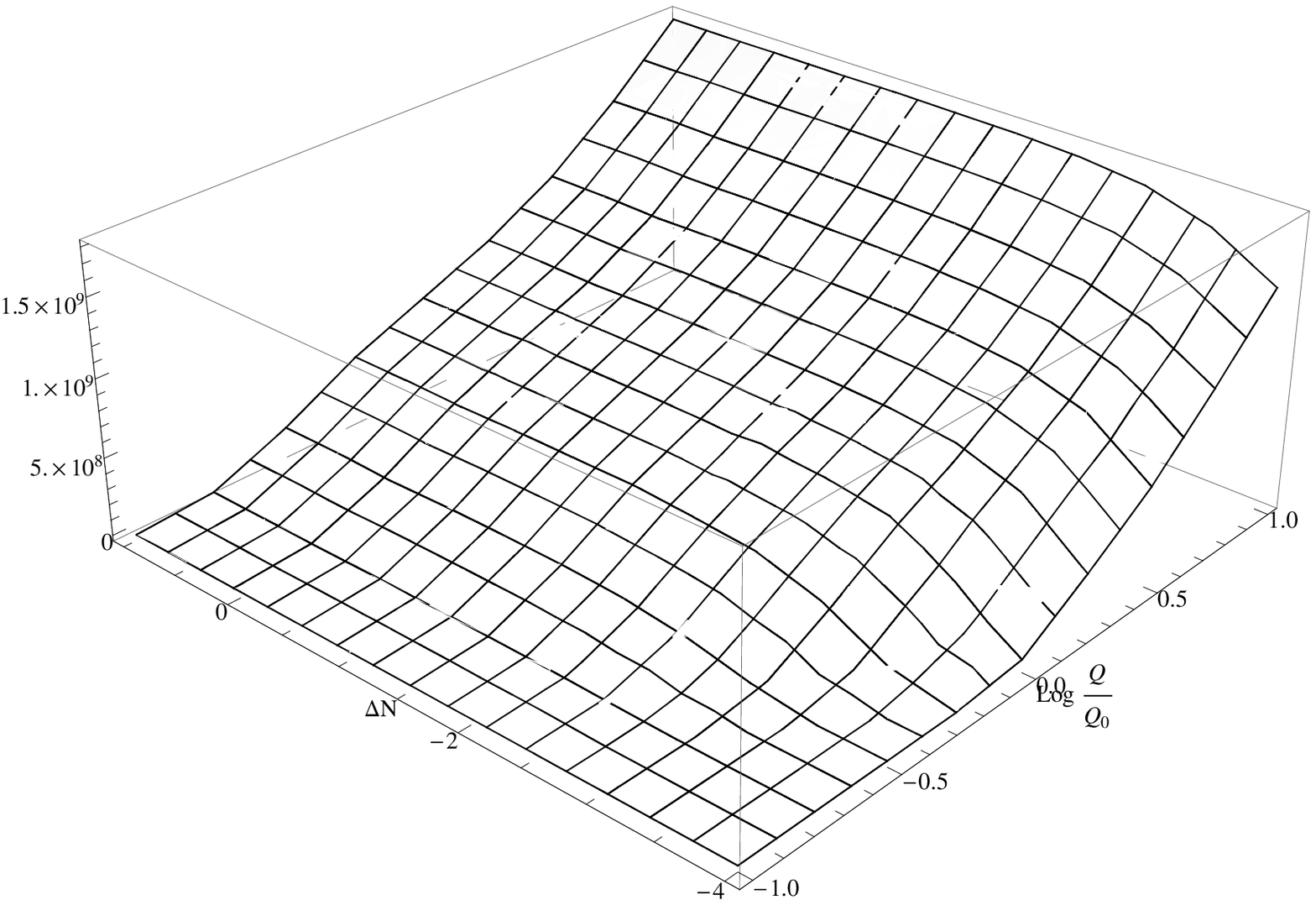}
}
\subfigure[]{
\includegraphics[width=2.5 in]{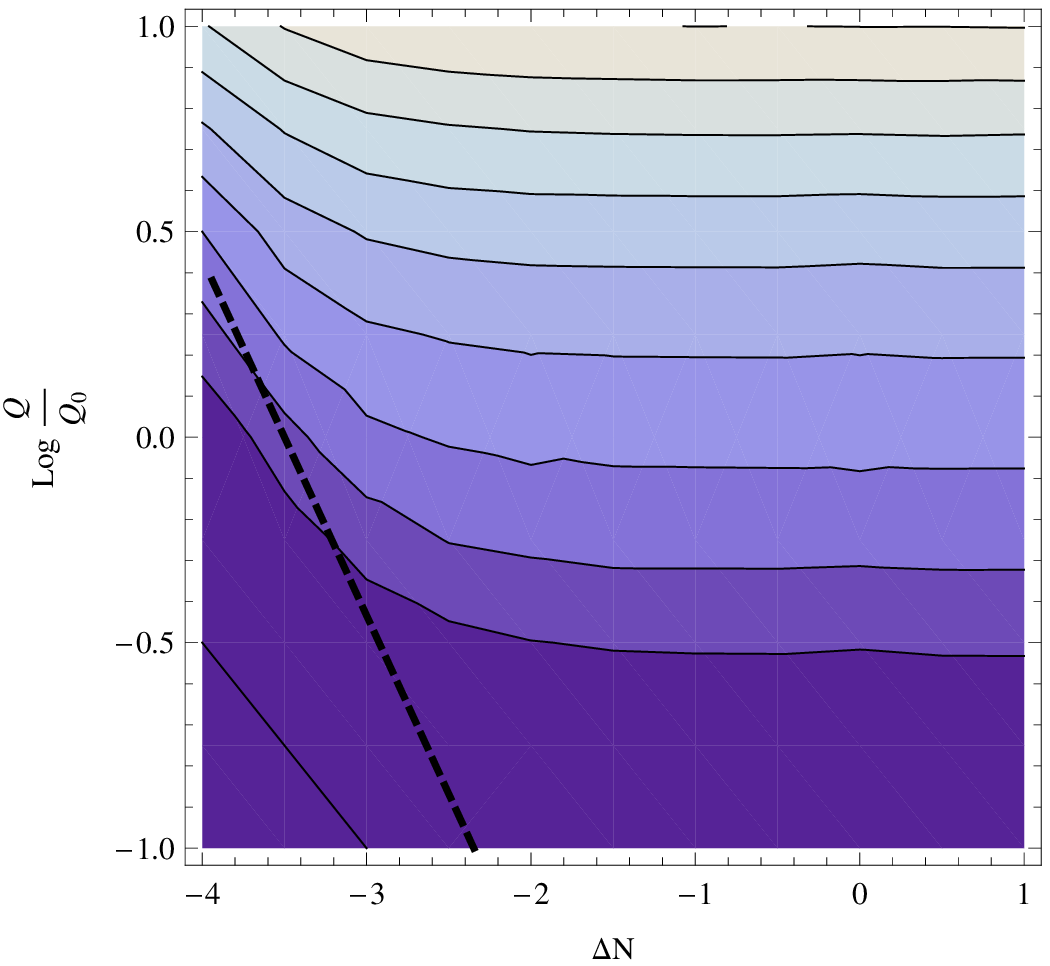}\label{fig:QNContourPlot}
}
\caption{The integrated SFR as a function of $\Delta N$ and $Q$, at fixed $\rho_\Lambda = \rho_{\Lambda_0}$~.  The contour plot (b) also shows the boundary $t_{\rm peak}(Q)>t_{\rm c}$ expected analytically, but this analytic boundary is not a good approximation to the true catastrophic boundary.
}
\label{fig:QNIntegratedPlots}
\end{figure}

If we also start to vary $Q$, the story gets only slightly more
complicated. Increasing $Q$ causes structure formation to happen
earlier, resulting in more star formation. In the multivariate
picture, then, it is helpful to think of $Q$ as changing the unsuppressed $t_{\rm
  peak}$.  For the unsuppressed case, the universe can be approximated as matter-dominated.  Looking at Eq.~\ref{eq:tpeak} and recalling that
$G\sim a\propto t^{2/3}$ in a matter-dominated universe, and
neglecting for the sake of computational ease the dependence on
$M_{\rm min}$, we see that $t_{\rm peak}\propto Q^{-3/2}$. Now we can
quantitatively assess the dependence of star formation on $Q$ and
$\rho_\Lambda$, for instance. We know that $t_\Lambda=t_{\rm peak}$
when $Q=Q_0$ and $\rho_\Lambda = 100\,\rho_{\Lambda_0}$, and this
condition should generally mark the boundary between star formation
and no star formation. Therefore we can deduce that the boundary in
the two-dimensional parameter space is the line $\log_{10}
\rho_\Lambda/\rho_{\Lambda_0} = 2 + 3 \log_{10} Q/Q_0$~. This line is shown
in Fig.~\ref{fig:LambdaQContourPlot}, where it provides a
good approximation to the failure of star formation.

\begin{figure}[h]
\center
\includegraphics[width=3.5 in]{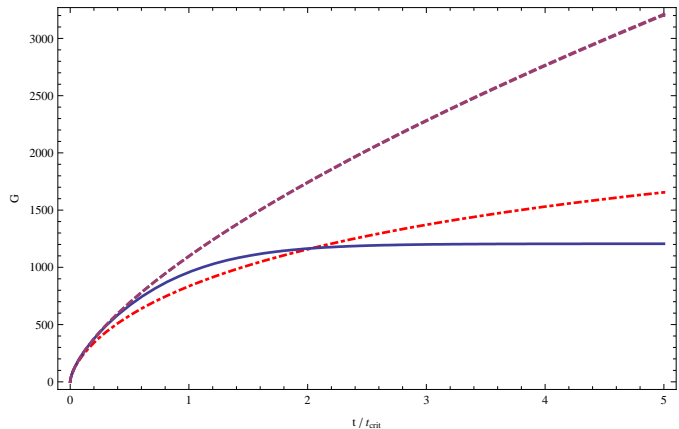}
\caption{The linear growth factor for three different universes: a flat $\rho_\Lambda\neq0$ universe (solid), a curved $\rho_\Lambda=0$ universe (dot-dashed), and a flat, $\rho_\Lambda=0$ (unsuppressed) universe (dshed).  The former two universes have been chosen so that $t_{\rm c} = t_\Lambda \equiv t_{\rm crit}$, and the time axis is given in units of $t_{\rm crit}$.  In the $\rho_\Lambda$ case, the growth factor can be well-approximated as being equal to the unsuppressed case up until $t=t_{\rm crit}$, and then abruptly becoming flat.  This approximation is manifestly worse for the curvature case.  In the curvature case, the growth factor starts to deviate from the unsuppressed case well before $t=t_{\rm c}$, and continues to grow well after $t=t_{\rm c}$.  This effect makes the case of strong curvature harder to approximate analytically.
}
\label{fig:GrowthFactorPlot}
\end{figure}

We can repeat the same analysis for the case of fixed cosmological constant with varying curvature and perturbation amplitude.  In Fig.~\ref{fig:QNIntegratedPlots} we show the integrated star formation rate as a function of $\Delta N$ and $Q$ for fixed $\rho_\Lambda=\rho_{\Lambda_0}$.  The dashed line in Fig.~\ref{fig:QNContourPlot} marks the boundary $t_{\rm c} = t_{\rm peak}$, where the time of curvature domination equals the time of the unsuppressed peak (again computed according to Eq.~\ref{eq:tpeak}).  In this case the line is given by $(2\Delta N +7)\log_{10} e = -\log_{10} Q/Q_0$, owing to the fact that $t_{\rm c}\propto\exp(3\Delta N)$.  Unlike the case of $\rho_\Lambda$, this line obviously does not mark the transition from structure formation to no structure formation.  The reason has to do with the details of the nature of the suppression coming from curvature versus that coming from the cosmological constant.  In the case of the cosmological constant, it is a very good approximation to say that structure formation proceeds as if there were no cosmological constant up until $t=t_\Lambda$, whereupon it stops suddenly.  Curvature is far more gradual: its effects begin well before $t_{\rm c}$ and structure formation continues well after $t_{\rm c}$.  Fig.~\ref{fig:GrowthFactorPlot} illustrates this point clearly.  There we see that while perturbations are suppressed in both the high curvature and large $\rho_\Lambda$ cases, it is only in the $\rho_\Lambda$ case that the naive approximation of a ``sudden end" to structure formation is valid.  For the curvature case there is no simple approximation, and the full model is necessary to give an accurate assessment of the situation.

\acknowledgments We thank K.~Nagamine for providing us with a
compilation of data points for the observed star formation rate. We
are grateful to C.~McKee, J.~Niemeyer, and E.~Quataert for
discussions.  We also thank J.~Carlson for collaboration at the early stages of this project.  This work was supported by the Berkeley Center for
Theoretical Physics, by a CAREER grant (award number 0349351) of the
National Science Foundation, by FQXi grant RFP2-08-06, and by the
U.S.\ Department of Energy under Contract DE-AC02-05CH11231.

\bibliographystyle{JHEP-2}
\bibliography{AstroBib}
\end{document}